\documentclass[times,authoryear,nopreprintline]{elsarticle}

\usepackage{amssymb}

\usepackage[figuresright]{rotating}

\usepackage{natbib}
\usepackage{amsfonts}		
\usepackage{booktabs}  
\usepackage{caption}	
\usepackage{makeidx}         
\usepackage{graphicx}
\usepackage{subfig}	        
\usepackage{multicol}        
\usepackage[bottom]{footmisc}
\usepackage{epstopdf}
\usepackage{floatrow}		
\newfloatcommand{capbtabbox}{table}[][\FBwidth]
\usepackage{multirow}
\usepackage{blindtext}
\makeindex             
\usepackage{comment}
\includecomment{comment}
\makeatother
\usepackage{breakcites}
\usepackage{amsmath}
\usepackage{array}
\usepackage{float}
\usepackage{algorithm}
\usepackage{algorithmic}
\usepackage{mathtools}
\usepackage{calc}  
\usepackage{enumitem}
\setlist[description]{leftmargin=\parindent,labelindent=\parindent}
\usepackage{graphicx}

\usepackage{nomencl}
\makenomenclature 

\usepackage{apalike}

\usepackage{color}
\usepackage{placeins} 

\long\def\ignore#1{}

\begin{document}

\begin{frontmatter}




\title{International Portfolio Optimisation with Integrated Currency Overlay Costs and Constraints}


\author{Nonthachote Chatsanga}
\ead{psxnc2@nottingham.ac.uk}

\author{Andrew J. Parkes}
\ead{Andrew.Parkes@nottingham.ac.uk}

\address{The Automated Scheduling, Optimisation and Planning (ASAP) Group, School of Computer Science, The University of Nottingham, Nottingham NG8 1BB, UK}

\begin{abstract}
Portfolio optimisation typically aims to provide an optimal allocation that minimises risk, at a given return target, by diversifying over different investments. However, the potential scope of such risk diversification can be limited if investments are concentrated in only one country, or more specifically one currency. Multi-currency portfolio is an alternative to achieve higher returns and more diversified portfolios but it requires a careful management of the entailed risks from changes in exchange rates.

The deviation between asset and currency exposures in a portfolio is defined as the ``currency overlay''.
This paper addresses risk mitigation by allowing currency overlay and asset allocation be optimised together. We propose a model of the international portfolio optimisation problem in which the currency overlay is constructed by holding foreign exchange rate forward contracts. Crucially, the cost of carry, transaction costs, and margin requirement of forward contracts are also taken into account in portfolio return calculation. This novel extension of previous overlay models improves the accuracy of risk and return calculation of portfolios; furthermore, our experimental results show that inclusion of such costs significantly changes the optimal decisions. Effects of constraints imposed to reduce transaction costs associated are examined and the empirical results show that risk-return compensation of portfolios varies significantly with different return targets. 
\end{abstract}


\end{frontmatter}


\section{Introduction}
\label{sec:intro}

A portfolio is a collection of assets that are different in terms of return and risk. In general, assets with higher return have higher risk (generally represented by standard deviation of returns from their expected values).
Nonetheless, when assets are held together, there presents diversification benefit from imperfect correlation which subsequently reduce the risk of portfolios. 
The key purpose on constructing a portfolio is hence to take the most advantage from asset correlations and achieve the lowest risk at the desirable level of return, \citep{Markowitz1952}.

To widen the scope of diversification rather than diversifying asset holdings across asset classes in only one single country, international investment is hence considered. 
Theoretically, investment in multiple countries could reduce exposure regarding systematic risk in specific markets and offer more opportunities to gain higher profits from promising foreign markets. 
The key advantage of multi-currency portfolios arises from potentially larger extent of risk diversification; as it is more likely that assets in similar economies are more correlated to their peers than those in different economies. \citep{Levy1970} document that low correlation between investment returns from developed and developing countries leads to the reduction of the portfolio variance. 

However, such benefits induce extra exposure to exchange rate variation which could raise or ruin the portfolio values at the same time. The importance of managing exchange rate risk is vital for business as addressed in \citep{Kim2009} and applied widely throughout various industries. \citep{Eun1988} discuss the effect of exchange rates on multi-currency portfolios and that, in adverse case, exchange rate losses could possibly override asset gains. They hence proposed a hedging strategy by short selling foreign currencies at forward rates. Their results show that such hedging strategy outperforms the unhedged one. Subsequently, \citep{Glen1993} employ forward contracts to hedge against foreign currency depreciation and formulated the optimisation problem that allows asset allocation and forward contract positions to be optimised together at the same time. 
They discovered that portfolios that are hedged limited to the size of foreign asset holdings perform better than portfolios employing the unitary hedging strategy\footnote{Also known as a fully-hedged portfolio which removes all exposure in foreign currencies to only exposure in local currency. For instance, if a USD-funded portfolio invests in US, UK and Japan, then the fully-hedged portfolio will have only currency exposure in USD.}.

Although there is a strong linkage between security markets and exchange rates much of the time they are not changing in the same direction. 
In order to manage currency exposure of international portfolios separately from asset allocation, the currency overlay is employed in devising hedged portfolios. 
In early development, currency hedging is applied to portfolios after obtaining optimal asset allocation, then the adjustment on initial currency exposure is ``overlaid'' on the portfolios, \citep{Nakakubo2004}. 
The most significant benefit of imposing currency overlay is that currency exposure management is separated from managing underlying assets. This flexibility becomes even more prominent when investing in a country whose asset markets negatively correlate to its exchange rates. The portfolio can be hedged so that has limited exposure to exchange rate risk but still maintains exposure in asset markets abroad. 
Besides, a well-managed currency overlay can also be used to manipulate currency positions for speculation on exchange rate markets which could enhance return to the portfolios, \citep{Levich1993}.

The primary and novel contribution of this paper is to develop optimisation model that includes a realistic and  comprehensive cost model of forward contracts. We develop a methodology to formulate the currency overlay by combining foreign exchange forwards and incorporating it into the multi-currency portfolio optimisation model. Crucially, we introduce an approach to include the cost of carry of forward contracts and associated transaction costs into the process of risk and return calculation which improves the accuracy in determining total returns and volatilities of portfolios. The importance of transaction costs are widely examined in financial and portfolio management literature. In essence, transaction costs affect the net gain to investments and actual performance of portfolios, \cite{Roll1984}. \citep{ArnottWagner1990} mention that ignoring the transaction costs would lead to inefficient portfolios. To our knowledge, the implementation of transaction costs on portfolio currency overlay and its effects on risk and return have not been investigated before in the literature. Another contribution is to study the effects of particular constraints imposed to reduce transaction costs from holding currency overlay. Since currency overlay utilisation is differed by composition of assets and currencies in a portfolio, different impacts from transaction costs reduction should affect the risk-return compensation of portfolios differently at each return target.

The benefits of the introduced model are twofold. Firstly, the resulting portfolios are optimal in both asset and currency points of view. With the currency overlay allowed, if adjustment on currency exposure provides no significant improvement on risk and return of a portfolio then the result is exactly the same as the one obtained with no currency overlay allowed. On the other hand, if in some circumstances that extra currency exposure helps improve risk-return reward of the portfolio, then the portfolio holds forward contracts to form the currency overlay. Secondly, by constructing currency overlay from a collection of foreign exchange forwards, it provides flexibility to impose different transaction costs on individual forwards which improves accuracy of optimal solutions after all.

The rest of the paper is organised as follows. The next section demonstrates how currency overlay is defined and how it is integrated into a portfolio construction process, followed by the approach demonstrating the inclusion of the cost of carry and transaction costs of forward contracts in the return and risk calculation and then formulating the optimisation model with associated currency overlay constraints. Section \ref{sec:res_disc} reports the experiment results and section \ref{sec:concl} provides the summary of this study.

\section{Methodology}
\label{sec:method}

\begin{figure}[tbhp]
\begin{centering}
\includegraphics[scale=0.39]{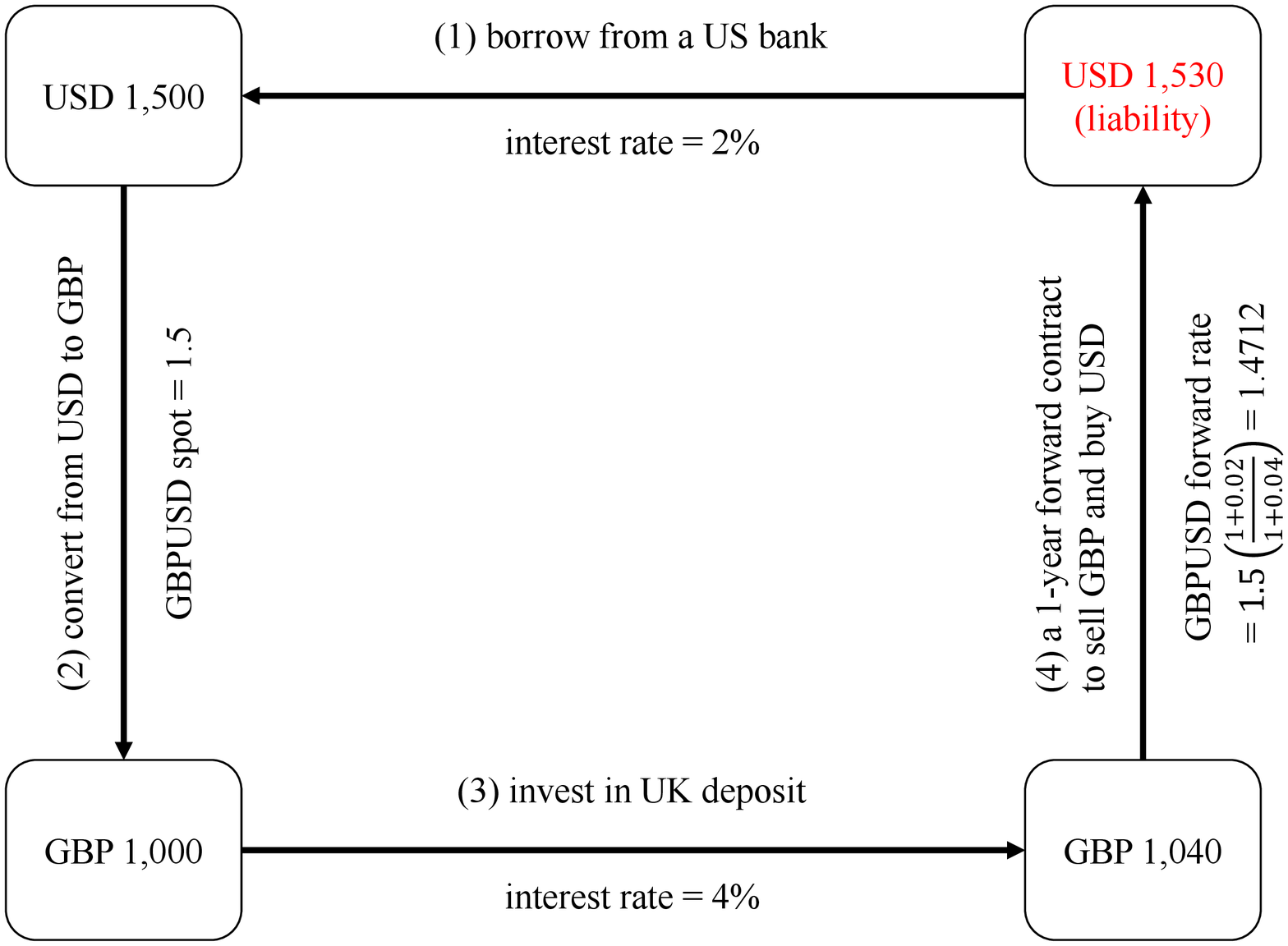}
\par\end{centering}
\caption{Illustration of how interest rates of two countries affect the corresponding forward exchange rate.}
\label{forward_CoC}
\end{figure}

When a portfolio invests in multiple currencies, there is currency risk entailed from exchange rates fluctuation. 
A wide range of financial derivatives can be used to mitigate currency risk such as foreign exchange forward contracts, foreign exchange futures and foreign exchange options. 
 
A foreign exchange forward contract is basically an obligation to buy or sell securities at future date with the price agreed today. 
The agreed price or a forward rate is slightly different from today's market price so as to prevents an investor from making a riskless profit by exploiting the interest rate differential.

The similar tool to forward contracts and that can be used to hedge foreign exchange rates risk is foreign exchange futures. 
The key difference is that a currency forward contract is a private over-the-counter transaction between counterparties known to each other, on terms agreed between themselves, while a currency futures contract is traded on a public exchange, e.g., the International Money Market (IMM) division of the Chicago Mercantile Exchange (CME). 
The futures contracts thus have standardised features such as units of trading, delivery and settlement dates and minimum price increments which could cause several constraints. Forward contracts are therefore preferred to model portfolio optimisation over futures contracts in terms of customisation flexibility.

Foreign exchange rate options is structured to serve the same purpose as forward contracts but with the main distinction that they offer the right to buy or sell at future date in contrast of obligation. 
As there is a choice not to exercise the right, there is a cost attached (details on options pricing can be found in \citep{Wilmott1993}) as opposed to forward contracts.  
Hence in terms of transaction cost, hedging a portfolio with forward contracts is more desirable.

In terms of valuation, when there is a change in the prices of underlying assets, gain and loss on forwards and options are impacted differently. 
Since valuation of futures and forwards positions are linear in the price of underlying securities, computation of gain and loss is therefore straightforward and similar to that of other assets in the portfolio. 
In contrast, for options their prices move in non-linear fashion with underlying assets so they need sophisticated predictions to calculate gain and loss. 
Therefore, in order to keep the portfolio valuation simple to study the benefit of currency overlay constraint, forward contracts are chosen as a hedging instrument in our study. 

\subsection{Cost of Carry of Foreign Exchange Forwards} 

The relationship between the spot rate and the forward rate is determined by the difference in the interest rates earned on the respective currency pairs, known as the ``cost of carry''. The idea is that buying forward contract is equivalent to buying an underlying asset now and pay the carry until the end of the contract. To exemplify how interest rates between two countries affect the forward rates and gain or loss on holding foreign exchange forwards, we consider Figure \ref{forward_CoC} for a US investor who:
\begin{itemize}
\item[1)] Borrows USD 1,500 from a US bank under the annual interest rate of 2\%. He is now obliged to pay back USD 1,530 to the bank one year later.
\item[2)] With the spot exchange rate of GBPUSD = 1.500, he then converts USD 1,500 to GBP 1,000.
\item[3)] He invests in a deposit in UK to earn a profit of 4\% per year for the sum of GBP 1,040. The final amount will be converted to USD one year later to pay the obligation to the US bank.
\item[4)] To avoid loss from exchange rates, he enters into a forward contract to sell GBP and buy USD in one year later.
\end{itemize}

We can see in step 4) that if the forward rate to sell GBP and buy USD in one year ahead is equal to the spot exchange rate in step 2), the investor can then make a round-tripping with the profit equal to the interest rate differential of 4\% - 2\% = 2\% per year. To eliminate such arbitrage opportunity, the 1-year forward rate for GBPUSD is priced as
\begin{equation}
\label{fwd_rates}
F_{GBPUSD} = S_{GBPUSD} \frac {(1+i_{GBP})} {(1+i_{USD})} 
\end{equation}
where $i_{GBP}$ and $i_{USD}$ are respectively the interest rates of UK and US, $S_{GBPUSD}$ is the spot exchange rate and $F_{GBPUSD}$ is the forward exchange rate of GBPUSD. More details on how the forward rates are priced are given in \citep{Korajczyk1985}.

The fair price of the forward rate in this example provides less amount of USD comparing to the one converted by the spot rate (1.4712 to 1.5000) to offset a profit from borrowing from low to invest in high interest rates. 
Equivalently, in this case, holding a foreign exchange forward contract incurs a cost of carry by 2\% - 4\% = -2\% which matches the profit from exploiting the interest rate differential. 
Therefore, the cost of carry from holding a foreign exchange forward contract can be computed by
\begin{equation}
\label{CoC}
\text{Cost of Carry} = i_{buy} - i_{sell}	
\end{equation}
where $i_{sell}$ is an interest rate of a country that one wants to sell the currency off so as to buy another currency and $i_{buy}$ is an interest rate of a country that one desires to buy.

\subsection{Currency Overlay} 

\begin{table}[tbhp]
\centering
\begin{tabular}{lccccccc}
\toprule 
 & \multicolumn{3}{c}{{\small{Hedged}}} &  & \multicolumn{3}{c}{{\small{Unhedged}}}\tabularnewline
\cmidrule{2-4} \cmidrule{6-8} 
 & {\small{US}} & {\small{UK}} & {\small{JP}} &  & {\small{US}} & UK & JP\tabularnewline
\midrule 
{\small{asset exposure (\%)}} & {\small{35}} & {\small{45}} & {\small{20}} &  & {\small{35}} & 45 & 20\tabularnewline
{\small{currency exposure (\%)}} & {\small{27}} & {\small{55}} & {\small{18}} &  & {\small{35}} & 45 & 20\tabularnewline
{\small{overlay position (\%)}} & {\small{-8}} & {\small{10}} & {\small{-2}} &  & \multicolumn{3}{c}{{\small{-}}}\tabularnewline
{\small{total overlay (\%)}} & \multicolumn{3}{c}{{\small{10}}} &  & \multicolumn{3}{c}{{\small{-}}}\tabularnewline
\bottomrule
\end{tabular}
\caption{Sample portfolios with and without currency overlay.}
\label{table:example}
\end{table}

Consider a portfolio that invests in different countries, basically, the value of a portfolio are affected by two sources of returns. One is from asset prices plus dividends or other interest-bearing incomes and another one is from gain or loss of exchange rates. Investment in each country is thus portrayed as a composition of exposure in asset markets and exposure in exchange rates. This structure also facilitates adjustment on currency exposure and hence dissipate risk foreign currency positions. The alteration made on currency exposure is defined as currency overlay which modifies the status-quo currency positions of unhedged portfolios. 

Clarification of how currency overlay helps manage exchange rate risk in multi-currency portfolios is given in Table \ref{table:example}. It shows a portfolio investing in three stock markets, the United States, the United Kingdom and Japan with an allocation of 35\%, 45\% and 20\% respectively. If the base currency of a portfolio is USD and a portfolio manager has a view that GBPUSD will appreciate while USDJPY will depreciate, he can hedge the portfolio with forward contracts so that the final currency exposure lies in his favour -- more holding in pound sterling and less exposure on Japanese yen. In contrast, if he has no view on exchange rate movements, the portfolio remains unhedged and there exists no overlay position. 

The given example treats choices of currency overlay as an ex-post decision which relies considerably on personal judgment and experience. To avoid personal discretion, there has been an improvement to incorporate hedging decisions into optimisation process, for instance, in the studies of  \citet{Adjaoute1995,Beltratti1999,Brown2012,Glen1993,LarsenJr2000,Rudolf1998,Topaloglou2002,Topaloglou2008}. The currency overlay positions in the past literature are, however, defined as a difference between asset and currency allocation and are not constructed from pairs of foreign exchange forward contracts. This limitation disallows implementation of different transaction costs on each forward pairs and imposition of constraints related to transaction costs reduction. Our approaches on the structure of portfolios and the inclusion of cost of carry and associated transaction costs of foreign exchange forwards are introduced to bridge those gaps which will make the optimisation model better accommodate practical constraints and deliver more accurate optimal solutions.

\subsubsection{Structure of a Portfolio with Overlay Constraints}

Since our currency overlay is built from a combination of foreign exchange forwards, there can be many choices possible to form a currency overlay from various number of forward pairs as shown in Table \ref{table:choice_overlay}. Note that every choice results in the same cost of carry as it creates the same overlay position.
In fact, for an investment of $C$ currencies, there exists at most ${C \choose 2}$ different forward contracts. To optimally choose forward pairs from all available choices, we introduce a new structure of currency overlay and incorporate it in the portfolio optimisation problem. Suppose that an investment plan is to invest in $A$ asset classes from $C$ countries, exposure of assets and currencies in a portfolio can be characterised as in Table \ref{table:overlay_structure} with the following notations:
\begin{description}[leftmargin=!,labelwidth=\widthof{\bfseries cccccc}]
\item[\textnormal{i}] Index of asset classes; $i=1, ..., A$
\item[\textnormal{j}] Index of countries, or synonymously currencies; $j=1, ..., C$
\item[\textnormal{k}] Index of forward contracts $k=1, ..., K$
\item[\textnormal{$a_{ij}$}] Exposure to asset class $i$ of country $j$
\item[\textnormal{$f_{kj}$}] Forwards position of contract $k$ on country (currency) $j$
\end{description}

The forward position $f_{kj}$ represents how much additional exposure is added into or taken off from a pair of currencies. Since each forward contract contains only a pair of currencies, given that there are $C$ countries to invest, then there are $K = {C \choose 2}$ different forward contracts in total. Considering exposure on each country $C$ from a forward contract $K$, since the exposure from a pair of currencies are equal when being valued in a portfolio's base currency, then it is strictly required that $\underset{j}\sum f_{kj} = 0$ for all $k=1,...,K$. 

\begin{table}[H]
\centering
\begin{tabular}{llccc}
\toprule
\multicolumn{2}{l}{} & {\small{USD}} & {\small{GBP}} & {\small{JPY}}\tabularnewline
\midrule
\multicolumn{2}{c}{overlay position (\%)} & {\small{-8}} & {\small{10}} & {\small{-2}}\tabularnewline[.1cm]
{\small{choice 1}} & {\small{USDJPY}} & {\small{2}} &  & {\small{-2}}\tabularnewline
 & {\small{GBPUSD}} & {\small{-10}} & {\small{10}} & \tabularnewline[.1cm]
{\small{choice 2}} & {\small{GBPJPY}} &  & {\small{2}} & {\small{-2}}\tabularnewline
 & {\small{GBPUSD}} & {\small{-8}} & {\small{8}} & \tabularnewline[.1cm]
{\small{choice 3}} & {\small{USDJPY}} & {\small{1}} &  & {\small{-1}}\tabularnewline
 & {\small{GBPUSD}} & {\small{-9}} & {\small{9}} & \tabularnewline
 & {\small{GBPJPY}} &  & {\small{1}} & {\small{-1}}\tabularnewline
\bottomrule
\end{tabular}
\caption{Sample choices to construct the same currency overlay.}
\label{table:choice_overlay}
\end{table}

Denoting $\textbf{f}_{k}=(f_{k1}, ..., f_{kC})$ a vector of exposure from a forward contract $k$, the previous requirement implies that only two elements of $\textbf{f}_{k}$ represent the exposure with one being equal to a negative value of another, while the rest of the elements only takes a value of zero. To avoid putting those requirements into constraints of an optimisation problem, we define $f_{kj}$ as an element of a matrix $\textbf{F}$ in which
\begin{equation}
\textbf{F}  \,\stackrel{\text{def}}{=}\, \textbf{T} \circ (\textbf{1}^{\textrm{T}} \otimes \textbf{q})
\label{Fmatrix}
\end{equation}
where $\circ$ is the Hadamard product operator, $\otimes$ is the Kronecker product operator, $\textbf{T}$ is a $K \times C$ combinatorial matrix of $\{-1,0,1\}$, $\textbf{1}$ is a $C \times 1$ column vector of ones and $\textbf{q}$ is a $K \times 1$ column vector determining the size of exposure.

\begin{table}[H]
\resizebox{\linewidth}{!}{
\centering
\begin{tabular}{lllllll}
\toprule
 &  & Country 1 & $\cdots$ & Country j & $\cdots$ & Country C\tabularnewline
\midrule
Asset class 1 &  & $a_{11}$ & $\cdots$ & $a_{1j}$ & $\cdots$ & $a_{1C}$\tabularnewline
$\vdots$ &  & $\vdots$ & $\ddots$ & $\vdots$ & $\ddots$ & $\vdots$\tabularnewline
Asset class $i$ &  & $a_{i1}$ & $\cdots$ & $a_{ij}$ & $\cdots$ & $a_{iC}$\tabularnewline
$\vdots$ &  & $\vdots$ & $\ddots$ & $\vdots$ & $\ddots$ & $\vdots$\tabularnewline
Asset class $A$ &  & $a_{A1}$ & $\cdots$ & $a_{Aj}$ & $\cdots$ & $a_{AC}$\\[.2cm]
\rule{0pt}{2.5ex}Forward position 1 &  & $f_{11}$ & $\cdots$ & $f_{1j}$ & $\cdots$ & $f_{1C}$\tabularnewline
$\vdots$ &  & $\vdots$ & $\ddots$ & $\vdots$ & $\ddots$ & $\vdots$\tabularnewline
Forward position $k$ &  & $f_{k1}$ & $\cdots$ & $f_{kj}$ & $\cdots$ & $f_{kC}$\tabularnewline
$\vdots$ &  & $\vdots$ & $\ddots$ & $\vdots$ & $\ddots$ & $\vdots$\tabularnewline
Forward position $K$ &  & $f_{K1}$ & $\cdots$ & $f_{Kj}$ & $\cdots$ & $f_{KC}$\\[.2cm]
\rule{0pt}{2.5ex}Asset exposure &  & $\sum\limits_{i=1}^{A}a_{i1}$ & $\cdots$ & $\sum\limits_{i=1}^{A}a_{ij}$ & $\cdots$ & $\sum\limits_{i=1}^{A}a_{iC}$\\[.2cm]
Overlay position &  & $\sum\limits_{k=1}^{K}f_{k1}$ & $\cdots$ & $\sum\limits_{k=1}^{K}f_{kj}$ & $\cdots$ & $\sum\limits_{k=1}^{K}f_{kC}$\\[.2cm]
Currency exposure &  & $\sum\limits_{i=1}^{A}a_{i1}+\sum\limits _{k=1}^{K}f_{k1}$ & $\cdots$ & $\sum\limits_{i=1}^{A}a_{ij}+\sum\limits _{k=1}^{K}f_{kj}$ & $\cdots$ & $\sum\limits_{i=1}^{A}a_{iC}+\sum\limits _{k=1}^{K}f_{kC}$\\[.2cm]
Total overlay &  & \multicolumn{5}{l}{$\frac{1}{2}\sum\limits_{j=1}^{C}\left|\sum\limits _{k=1}^{K}f_{kj}\right|$}\tabularnewline
\bottomrule
\end{tabular} 
}
\caption{Structure of multi-currency portfolios.}
\label{table:overlay_structure}
\end{table}

The combinatorial matrix $\textbf{T}$ of ternary variables is constructed by first specifying $\textbf{T}$ as a matrix of zeros of size $K \times C$, then we denote a set $\textbf{D}$ containing all combinatoric pairs from ${C \choose 2}$. At each row of $\textbf{T}$, the first member of each pair in $\textbf{D}$ specifies which element to take the value of 1 and the second member of the pair specifies the element that take value of -1. For example, if there are 4 countries to invest, that is, $C=4$, then $K = {4 \choose 2} = 6$ and so:
\begin{equation}
\textbf{D} = \{(1,2),(1,3),(1,4),(2,3),(2,4),(3,4)\}.
\nonumber
\end{equation} 
The matrix $\textbf{T}$ whose elements are specified according to $\textbf{D}$ is
\begin{equation}
\textbf{T} = 
\begin{bmatrix}
1  & -1 & 0  & 0 \\
1  & 0  & -1 & 0 \\
1  & 0  & 0  & 1 \\
0  & 1  & 1  & 0 \\
0  & 1  & 0  & 1 \\
0  & 0  & 1  & 1 
\end{bmatrix}.
\label{matrixT}
\end{equation}
Consequently, referring to (\ref{Fmatrix}), $\textbf{F}$ is equal to
\begin{align*}
\textbf{F} 
&=
\begin{bmatrix}
1  & -1 & 0  & 0 \\
1  & 0  & -1 & 0 \\
1  & 0  & 0  & -1 \\
0  & 1  & -1  & 0 \\
0  & 1  & 0  & -1 \\
0  & 0  & 1  & -1 
\end{bmatrix}
\circ
\left( 
\begin{bmatrix}
1  &  1 &  1 &  1 
\end{bmatrix}
\otimes
\begin{bmatrix}
q_1 \\ q_2 \\ q_3 \\ q_4 \\ q_5 \\ q_6
\end{bmatrix}
\right) \\
&=
\begin{bmatrix}
q_1  & -q_1 & 0  & 0 \\
q_2  & 0  & -q_2 & 0 \\
q_3  & 0  & 0  & -q_3 \\
0  & q_4  & -q_4  & 0 \\
0  & q_5  & 0  & -q_5 \\
0  & 0  & q_6  & -q_6 
\end{bmatrix} 
\end{align*}
and $f_{kj}$ is defined accordingly as an element of $\textbf{F}$. Subsequently, all requirements regarding the characteristics of forward contracts are completely fulfilled. 

With all necessary variables being defined, the constraints associated with overlay positions and forward contracts can be of the following examples:
\begin{enumerate}
\item \textit{Limited total overlay positions} - total overlay specifies how much currency exposure can deviate from asset exposure. Let $V$ denote the total overlay of a portfolio and $V_u$ the total overlay limit allowed on a portfolio, the constraint is defined as
\begin{equation}
V \,\leq\, V_u
\end{equation}
where $V=\dfrac{1}{2}\sum\limits_{j=1}^{C}\left|\sum\limits_{k=1}^{K} f_{kj}\right|$. Notice that $V_u=1$ allows total currency exposure deviate from total asset exposure up to 100\% of the portfolio while $V=0$ implies an unhedged portfolio which disallows any use of forward contract. More room for currency and asset exposures deviation provides more opportunity to shift from less-performing to better-performing currencies which hence better improve the risk-return profile of the portfolio.
\item \textit{Bounded currency exposure} - this constraint is imposed, for example, when the net short currency positions are not permitted on specific countries. the constraint can be used in tandem with restrictions of short positions on assets. As a consequence, a specific bound $E_j^{l}$ and $E_j^{u}$ of currency exposure can be imposed on each country. The constraint is thus specified as
\begin{align}
E_j^{l} \leq \sum\limits_{i=1}^{A}a_{ij}+\sum\limits_{k=1}^{K}f_{{kj}}  \leq  E_j^{u};\quad  j = 1,..,C.
\end{align}
\item \textit{Limited number of forward contracts} - this can be viewed as a cardinality constraint on the number of forward contracts. In practice, more number of forward contracts means more operational burden which can be lessened by specifying a limit on number of contracts allowed. The constraint can be formulated as 
\begin{align}
l_k b_k  \leq q_k  \leq  u_k b_k; & \quad  k = 1,..,K, \\
\sum\limits_{k=1}^{K}b_k  \leq  G, & \\
b_k  \in \{0, 1\},
\end{align}
where $l_k$ and $u_k$ are respectively lower bound and upper bound for exposure sizes and $G$ specifies the total allowance on number of forward contracts.
\end{enumerate}

It is worth noting that the minimum contract size of forward contracts is customisable with counterparties, the constraint like minimum holding size is hence unnecessary.

\subsubsection{Incorporating Cost of Carry to Risk and Return Calculation of Portfolios}

We mention earlier that entering into forward contracts incurs the cost of carry which can be positive or negative depending on interest rate differential. Consider a portfolio holding three foreign exchange forwards as given in Table~\ref{tab:CoC}.
The cost of carry of each forward contract depends on which currency to sell or buy, corresponding interest rates, and position taken on the portfolio. For instance, selling JPY for USD at 1\% of the portfolio amounts the positive carry of $1\%\times2\%-1\%\times1\% = 0.01\%$ to the portfolio. Selling GBP for JPY, however, generates the negative carry of $-2\%\times4\%+2\%\times1\% = -0.06\%$ as a result of shifting exposure from country with high interest rate to the country with lower interest rate. The total overlay position is 8\% of the portfolio bearing the positive carry of 0.13\% from the three forward contracts combined. This amount of carry is added to the total return of the portfolio.

\begin{table}[bthp]
  \centering
  \caption{Costs of carry associated with foreign exchange rate forward contracts.}
    \begin{tabular}{rcccc}
    \toprule
          & \small{USD}   & \small{GBP}   & \small{JPY}   & {\small Cost of Carry} \\
    \midrule
    \small{interest rate (\%)} & {\small 2}   & {\small 4}   & {\small 1}   &  \\
    \rule{0pt}{2.5ex}{\small sell JPY, buy USD (\%)} & {\small 1}   &       & {\small -1}  & {\small 0.01} \\
    {\small sell USD, buy GBP (\%)} & {\small -9}  & {\small 9}   &       & {\small 0.18} \\
    {\small sell GBP, buy JPY (\%)} &       & {\small -2}  & {\small 2}   & {\small -0.06} \\
    \rule{0pt}{2.5ex}{\small overlay (\%)} & {\small -8}  & {\small 7}   & {\small 1}   & {\small 0.13} \\
    \bottomrule
    \end{tabular}
  \label{tab:CoC}
\end{table}

From Table \ref{tab:CoC}, the net cost of carry is in fact the product of interest rates and overlay positions. For an investment in any country $j$, the total return contributes to the portfolio is
\begin{equation}
\label{ret_j}
r_j = a_{j}r^a_j + c_{j}r^c_j + v_{j}i_j
\end{equation}
where $r_j$ is total return from investment in country $j$; $a_j$, $c_j$ and $v_j$ are respectively asset exposure, currency exposure and overlay position on country $j$; $r^a_j$, $r^c_j$ and $i_j$ are respectively expected asset return, expected currency return and expected interest rate of country $j$.

Since overlay position is defined as the difference in currency and asset exposures, equation (\ref{ret_j}) can be equivalently expressed as
\begin{align}
r_j &= a_{j}r^a_j + c_{j}r^c_j + (c_{j}-a_{j})i_j \nonumber \\
    &= a_{j}(r^a_j - i_j) + c_{j}(r^c_j + i_j).
\label{adjusted_return}
\end{align}
We define $r^a_j - i_j$ and $r^c_j + i_j$ as adjusted return of asset and adjusted return of currency, respectively. Equation (\ref{adjusted_return}) demonstrates that the portfolio total return (return from assets, currencies and costs of carry of foreign exchange forwards) is equal to the product of adjusted returns, asset exposure and currency exposure. This implies that the expression of overlay positions is not explicitly required to calculate total returns of a portfolio. In addition, if a portfolio holds no forward contract, the interest rate terms in equation (\ref{adjusted_return}) will be cancelled out, showing that the formulation in equation (\ref{adjusted_return}) generalises total return calculation of international portfolios.

Similarly to asset and currency returns, interest rates are not constant over time, volatility of interest rates is thus needed to be included in the calculation of portfolio risk. In accordance with return calculation of international portfolios, we apply equation (\ref{adjusted_return}) to adjust return time series for variance and covariance calculation. For example, denotes the covariance between S$\&$P500 index and GBPUSD by $\omega$, the adjusted-return covariance is calculated by
\begin{equation}
\label{adjusted_covar}
\omega = \text{cov}\left((X-Z_{US}),(Y+Z_{UK})\right)
\end{equation}
where $\text{cov}(\cdot)$ is a covariance function, $X$ is the time series of S$\&$P500 index returns, $Y$ is the time series of GBPUSD returns, $Z_{US}$ is the time series of US short-term interest rates and $Z_{UK}$ is the time series of UK short-term interest rates.

\subsubsection{Incorporating Transaction Costs of Forward Contracts}

Generally, transaction costs associated to holding forward contracts are categorised as fixed operating cost, bid-ask spread\footnote{The spread represents the difference between the highest price that a buyer is willing to pay (bid) for a security and the lowest price that a seller is willing to accept for it (ask). It is considered a cost as a security is purchased at an ask price while it is valued at a bid price which is always lower. To exemplify, suppose that stock A is quoted as bid = \$5.0 and ask = \$5.5 per unit (bid-ask spread = \$0.5), if an investor buys 100 units of stock A, he needs to pay \$550 for \$500 worth of stock A. Thus the wider the spread, the higher one needs to pay to acquire a security and the lesser the net profit.} and opportunity cost from margin requirement. All the costs involve in portfolio return calculation by means of a cost function reducing expected returns of portfolios. The optimisation problem can still be cast as single-period as the initial portfolio is the one without currency overlay implemented.

The fixed cost could be attributed to operational overheads charged from entering into a contract and maintaining the position until maturity date, the cost thus occurs when there is a forward position and does not proportionate with exposure. Bid-ask spreads act like break-even costs on holding securities and are attached to sizes of transaction, thus larger positions carry more costs on portfolios. Basically, determinants of bid-ask spreads are volatility and liquidity of respective forward pairs, thus each foreign exchange forward pair carries different charge unlike the fixed costs. Bid-ask spreads and fixed costs can be modelled as a cost function as follows,

\begin{equation}
\phi_{k}(q_k) = \alpha + \beta_{k} \left| q_k \right| \nonumber
\end{equation}
where $\phi(\cdot)$ is a transaction cost function, $\alpha$ is a fixed operating cost per contract, $\beta_k$ is a percent spread of each forward pair calculated from $(\textrm{ask price}-\textrm{bid price})/\textrm{ask price}$ and $q_k$ is exposure on each forward contract as defined by equation (\ref{Fmatrix}). Note that, in practice, buying and selling securities experience unequal percent spread but for liquid instruments like exchange rate forwards, the difference in buying and selling percent spreads could be insignificant, therefore we assume that both buying or selling take the same costs. In addition, the cost function is non-convex due to the presence of fixed operating costs.

The last component of transaction costs associated to forward contracts is considered an opportunity cost on account of margin requirement. Since the contract is marked to market daily, cash or liquid assets of values equivalent to some percentage of contract values must be reserved to ensure that all contract participants are able to meet the claims from continuous settlement process, \citep{Jarrow1981}. Because forwards are not traded on exchanges, the contract terms are not standardised and thus the percentages required on margin vary by counterparties. The existence of maintenance margin stipulates a portfolio to set aside its portion to cash and earn zero or nearly-zero return which diminishes attractiveness of holding forwards in portfolios. The relationship between cash portion and margin requirement is expressed by the following equation,

\begin{equation}
a_{mn} = M\sum\limits_{k=1}^{K}\left| q_k \right| \nonumber
\end{equation}
where $a_{mn}$ represents cash allocation in a portfolio, $M$ is a percentage of margin requirement and $q_k$ is exposure on each forward contract.

\subsubsection{Optimisation Problem with Overlay Constraints}
Based on the aforementioned overlay constraints and the portfolio structure as in Table \ref{table:overlay_structure}, the mean-variance portfolio optimisation problem can be set up with following additional notations,
\begin{description}[leftmargin=!,labelwidth=\widthof{\bfseries cccccc}]
\item[a] Vector of asset exposure; $\textbf{a} = [a_{11}, ..., a_{ij}, ..., a_{AC}]^{\textrm{T}}$.
\item[c] Vector of currency exposure; $\textbf{c} = [c_1, ..., c_j, ..., c_C]^{\textrm{T}}$ where \\ $c_j = \sum\limits_{i=1}^{A}a_{ij}+\sum\limits_{k=1}^{K}f_{{kj}}$; $j=1, ...,C$.
\item[x] Vector of decision variables; $\textbf{x} = [\textbf{a},\textbf{c}]^{\textrm{T}}$.
\item[r] Vector of expected returns; $\textbf{r} \in \mathcal{R}^{C(A+1)}$.
\item[\textnormal{$\mu$}] Target return of a portfolio.
\item[$\Omega$] Variance-covariance matrix of asset and currency returns.
\end{description}
Note that $\textbf{f}$ is a function of $\textbf{q}$ as in equation (\ref{Fmatrix}), thus the total number of decision variables is $AC + K$ and $\textbf{x} \in \mathcal{R}^{AC+K}$. The vector  $\textbf{r}$ contains adjusted expected returns of assets and currencies according to equation (\ref{adjusted_return}) by subtracting expected interest rates from asset returns and adding expected interest rates to currency returns. Each element of the covariance matrix $\Omega$ is calculated from adjusted return time series as exhibited in equation (\ref{adjusted_covar}). The mean-variance portfolio optimisation problem with overlay constraints is subsequently formulated as:

\begin{subequations}
\label{MVproblem}
\begin{align}
	\text{minimise}
       & \quad \textbf{x}^{\textrm{T}}\Omega\textbf{x} \\
    \text{subject to} 
       \label{port_ret}
	   & \quad \textbf{x}^{\textrm{T}}\textbf{r} - \sum\limits_{k=1}^{K}\phi_{k}(q_k) = \mu, \\
	   \label{1st_ovlcon}
	   & \quad \textbf{x} = [\textbf{a},\textbf{c}]^{\textrm{T}}, \\ 
	   & \quad \textbf{F} = \textbf{T} \circ (\textbf{1}^{\textrm{T}} \otimes \textbf{q}), \\
	   & \quad c_j = \sum\limits_{i=1}^{A}a_{ij}+\sum\limits_{k=1}^{K}f_{{kj}}; f_{{kj}} = \textbf{F}_{kj}, \\
	   \label{totovl}	   
	   & \quad \dfrac{1}{2}\sum\limits_{j=1}^{C}\left|\sum\limits_{k=1}^{K} f_{kj}\right| \,\leq\, V_u, \\	
	   \label{boundedccyexp}  
	   & \quad E_j^{l} \leq \sum\limits_{i=1}^{A}a_{ij}+\sum\limits_{k=1}^{K}f_{{kj}} \leq  E_j^{u}, \\	 
	   \label{1st_carcon}
	   & \quad l_k b_k \leq q_k  \leq  u_k b_k, \\
	   \label{last_carcon}
 	   & \quad \sum\limits_{k=1}^{K}b_k  \leq  G, \\
 	   \label{cashcon}
 	   & \quad a_{mn} = M\sum\limits_{k=1}^{K}\left| q_k \right|, \\
 	   & \quad \textbf{1}^{\textrm{T}}\textbf{a} = 1, \\
 	   & \quad \textbf{1}^{\textrm{T}}\textbf{c} = 1, \\
 	   & \quad 0 \leq a_{ij} \leq 1, \\ 	   
	   & \quad b_k  \in  \{0, 1\}.	\label{last_ovlcon}   
\end{align}
\end{subequations}

\section{Results and Discussions}
\label{sec:res_disc}

This section provides details on data sets used in the experiments, studies the impact of how constrained overlay influence return and risk of portfolios. Portfolios in the tests are constructed based on the formulation given in earlier section. We set USD as the base currency of a portfolio, meaning that we try to maximise profit or minimise risk in only USD. 

Portfolio investment is scoped to four major countries with four major currencies; the United States (USD), Germany (EUR), the United Kingdom (GBP) and Japan (JPY). The selection of major currencies ensure that foreign exchange forward contracts are available for all possible currency crosses. The number of asset classes are set as three, i.e., bonds and stocks to represent low and high risk investments and operating cash to cover margin requirement of forward contracts. With this setting, there comprises of 14 decision variables from $AC+K = (2\times4)+6$ where $A=2$ (number of asset classes), $C=4$ (number of currencies) and $K={4\choose2}=6$ (number of all distinct forward contracts).

\subsection{Data}

Government bond returns are collected from MorganMarkets by J.P. Morgan while stock and currency returns are retrieved from Bloomberg. Data frequency is monthly, spanning from Jan-00 to Jun-12. The interest rate of each country is proxied by its yield to maturity of 1-month treasury bill to reflect the risk-free return over a month. The yields to maturity of treasury bills are retrieved from Bloomberg. 

Expected returns of assets in each country are adjusted by subtracting with corresponding expected interest rate while an expected return of each currency is added with an expected interest rate of that country. Expected returns of assets, exchange rates are calculated using average historical returns. For expected interest rates, since the study period covers severe financial crisis in 2008 that forced central banks in large economies to keep interest rates historically low, the historical averages over Jan-00 to Jun-12 of yield to maturity are thus implausible to reflect the forward-looking expected interest rates. We therefore use average yields to maturity over Jun-11 to Jun-12 to represent recent information on expected interest rates. The estimated values are shown in Table \ref{tab:YTM}.

\begin{table}[H]
  \small
  \centering
  \caption{Average yields to maturity (as a proxy of interest rate) of 1-year treasury bills over Jun-11 to Jun-12. Note that the yield to maturity is generally given in terms of Annual Percentage Rate (A.P.R.), the figures shown are therefore converted to monthly rates by dividing the annual yields by 12 so that the resulting yields to maturity are comparable with other monthly returns.}
    \begin{tabular}{rcccc}
    \toprule
          & \multicolumn{4}{c}{average yield to maturity (\% monthly)} \\    
          \cmidrule{2-5}
          & US    & Germany & UK    & Japan \\
    \midrule
    \multicolumn{1}{c}{1-month treasury bill} & 0.004  & 0.012  & 0.053  & 0.008 \\
    \bottomrule
    \end{tabular}%
  \label{tab:YTM}%
\end{table}%

\begin{figure}[h]
\begin{centering}
\includegraphics[scale=0.55]{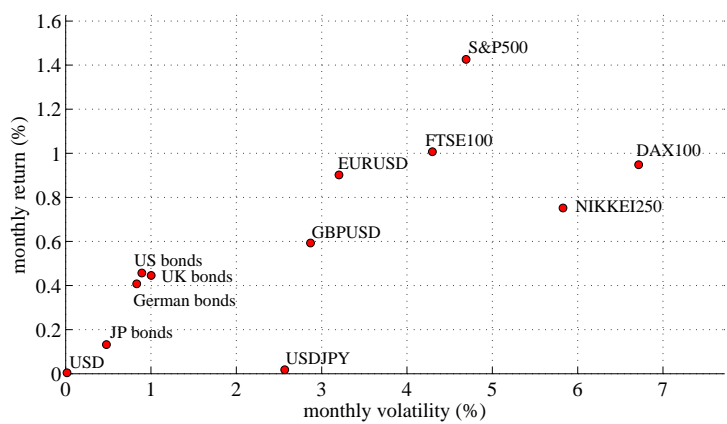}
\par\end{centering}
\caption{Adjusted expected returns and volatilities of assets and currencies.}
\label{RiskReturn}
\end{figure}

The covariance matrix is constructed from adjusted time series of asset returns and exchange rates as stated in equation (\ref{adjusted_covar}). The adjusted expected returns and volatilities are shown in Table \ref{tab:adjret}. The plot of risk (volatility) versus return of each security in the portfolio is illustrated in Figure \ref{RiskReturn}.

\begin{table}[htbp]
  \resizebox{\linewidth}{!}{
  \centering
  \caption{Adjusted expected returns used in the optimisation problems. Note that asset returns are in local currencies and exchange rate returns are measured against USD. The adjusted return of 0.004\% monthly on USD exchange rate is therefore solely from the expected US interest rate presented in Table \ref{tab:YTM}.}
    \begin{tabular}{rcccccccc}
    \toprule
          & \multicolumn{4}{c}{adjusted expected return (\% monthly)} & \multicolumn{4}{c}{adjusted volatility (\% monthly)} \\
    \cmidrule{2-5} \cmidrule{6-9}
          & US    & Germany & UK    & Japan & US    & Germany & UK    & Japan \\
    \midrule    
    government bond & 0.446 & 0.408 & 0.457 & 0.132 & 1.003 & 0.833 & 0.893 & 0.478 \\
    stock index & 1.426 & 0.948 & 1.007 & 0.752 & 4.692 & 6.714 & 4.297 & 5.828 \\
    exchange rate (against USD) & 0.004 & 0.902 & 0.593 & 0.018 & 0.017 & 3.203 & 2.870 & 2.567 \\
    \bottomrule
    \end{tabular}%
    \label{tab:adjret}%
    }  
\end{table}%

For the transaction costs of forward contracts, the fixed operating cost of each forward contract is arbitrarily set as 0.0001\% of portfolio value. The margin requirement of each forward contract is set as 10\% of the position size (the margin requirement of forward contracts is generally charged around 5 to 10\% on contract values, \citep{Levi2009}). The percent spreads of six possible forward pairs are collected from Bloomberg and are shown in Table \ref{tab:spread}. 

\begin{table}[H]
  \small
  \centering
  \caption{Average percent spreads, $(\textrm{ask price}-\textrm{bid price})/\textrm{ask price}$, of 1-month foreign exchange forward contracts during Jan-00 to Jun-12.}
    \begin{tabular}{lc}
    \toprule
          & average percent spread \\
          & (\% of ask price) \\ 
    \midrule
    USDEUR & 0.0036 \\
    USDGBP & 0.0051 \\
    USDJPY & 0.0050 \\
    EURGBP & 0.0042 \\
    EURJPY & 0.0068 \\
    GBPJPY & 0.0122 \\
    \bottomrule
    \end{tabular}%
  \label{tab:spread}%
\end{table}%

The percent spreads are calculated from $(\textrm{ask price}-\textrm{bid price})/\textrm{ask price}$. 
The values shown in Table \ref{tab:spread} are the averages over the period of Jan-00 to Jun-12. 
The percent spreads are applied on the forward positions selected by portfolios. 
Thus, increasing the number of forward pairs and larger overlay positions can generate more cost and lower return to portfolios. 
Since buying and selling forward contracts encounter indistinguishable percent spreads, we assume that all transactions in the experiments hereafter are subject to common bid-ask spreads shown in Table \ref{tab:spread}. 

\subsection{Experimental Studies}

We perform various experiments to investigate the benefit of incorporating currency overlay to international portfolios and effects of overlay constraints imposition on risk and return of portfolios. The first study exhibits a comparison on different approaches that currency overlay is implemented. The second experiment demonstrates how currency hedging affect the risk and return of portfolios. The third experiment focuses on the impact of margin requirement of foreign exchange forwards on risk-return compensation of portfolios. Then we study approaches to lessen transaction costs on overlay construction. 

Generally, smaller currency overlay positions reduce transaction costs but at the same time causes a deterioration in portfolio efficiency. We therefore aim to explore if there is any situation that saving transaction costs does not much affect risk-return reward of portfolios. The fourth experiment hence inspects if different total overlay limits affects risk of portfolios similarly at each return target, then the last experiment examines further how limiting maximum numbers of forward contracts allowed (which is one way to reduce the fixed-cost) in creating currency overlay could impact risk-return profile of portfolios.

The structure of our international portfolio is presented in Table \ref{tab:port_structure}. Note that the forward positions are defined as in equation (\ref{Fmatrix}), therefore they are expressed in terms of $q$. For each forward contract, a minus sign indicates selling and a plus sign indicates buying. For instance, the forward contract 1 (USDEUR) represents entering into a contract to sell EUR for USD at a proportion of $q_1$ of a portfolio. If the value of $q_1$ is negative, the contract is made reverse to sell USD for EUR instead. Since the portfolio is funded in USD, the operating cash is reserved only in USD and $a_{mn}$ in the constraint (\ref{cashcon}) is thus $a_{31}$. In consequence, the decision variables are a vector of 14 elements, i.e., $(a_{11}, \dots, a_{24}, q_1, \dots, q_6)$. 

Note that, however, the calculation of return and variance of a portfolio requires a vector of asset exposure and currency exposures (the vector $\textbf{x}$ defined in equation (\ref{1st_ovlcon})). The vector of asset and currency exposures is thus comprises of the first three rows (asset classes 1, 2 and 3) and the last row (currency exposure) of Table \ref{tab:port_structure}.

\begin{table}[tbhp]
  \resizebox{\linewidth}{!}{
  \centering
  \caption{The structure of international portfolio investing in four countries with two asset classes. All the variables $(a_{11}, \dots, a_{31}, q_1, \dots, q_6)$ are decision variables in the portfolio optimisation problem.}
    \begin{tabular}{lllll}
    \toprule
          & US    & Germany & UK    & Japan \\
    \midrule
    asset class 1 (bond) & $a_{11}$ & $a_{12}$ & $a_{13}$ & $a_{14}$ \\
    asset class 2 (equity) & $a_{21}$ & $a_{22}$ & $a_{23}$ & $a_{24}$ \\
    asset class 3 (cash) & $a_{31}$ &  &  &  \\
    \rule{0pt}{2.5ex}forward 1 (USDEUR) & $q_1$ & $-q_1$ &       &  \\
    forward 2 (USDGBP) & $q_2$ &       & $-q_2$ &  \\
    forward 3 (USDJPY) & $q_3$ &       &       & $-q_3$ \\
    forward 4 (EURGBP) &       & $q_4$ & $-q_4$ &  \\
    forward 5 (EURJPY) &       & $q_5$ &       & $-q_5$ \\
    forward 6 (GBPJPY) &       &       & $q_6$ & $-q_6$ \\
    \rule{0pt}{2.5ex}asset exposure & $a_{11}+a_{21}$ & $a_{12}+a_{22}$ & $a_{13}+a_{23}$ & $a_{14}+a_{24}$ \\
    overlay position & $q_1+q_2+q_3$ & $-q_1+q_4+q_5$ & $-q_2-q_4+q_6$ & $-q_3-q_5-q_6$ \\
    \rule{0pt}{2.5ex}\multirow{2}{*}{currency exposure} & $a_{11}+a_{21}+$ & $a_{12}+a_{22}-$ & $a_{13}+a_{23}-$ & $a_{14}+a_{24}-$ \\
    & $q_1+q_2+q_3$ & $q_1+q_4+q_5$ & $q_2-q_4+q_6$ & $q_3-q_5-q_6$ \\        
    \bottomrule
    \end{tabular}%
    \label{tab:port_structure}
  }  
\end{table}%

For associated constraints, the key parameters are an overlay limit $V$, lower bound and upper bound of currency exposure $E_j^l$ and $E_j^u$, lower bound and upper bound of forward positions $l_k$ and $u_k$ and a number of forward contracts allowed $G$. The constraint parameters in the experiments are imposed as in Table \ref{tab:constr_tab} if they are not stated otherwise. For other components of transaction costs, the bid-ask spreads are fixed at market averages as in Table \ref{tab:spread} and the fixed operating cost per forward contract is set at 0.0001\% of portfolio value.

\begin{table}
  \resizebox{\linewidth}{!}{
  \footnotesize
  \centering
  \caption{Parameter of constraints associated to currency overlay in international portfolio optimisation problem. In some experiments, some parameters are varied from these default values to study their effects on portfolios.}
    \begin{tabular}{llp{5cm}}
    \toprule
    Constraint & Parameter & Description \\
    \midrule
    constraint (\ref{totovl}) & $V_u=100\%$ & Total currency exposure can deviate from total asset exposure up to 100\% of a portfolio. \\
    constraint (\ref{boundedccyexp}) & $E_j^{l}=0$ and $E_j^{u}=1$ & Currency exposure of each country is always positive. \\
    constraint (\ref{1st_carcon}) & $l_k = -1$ and $u_k = 1$ & Upper and lower bounds of each forwards position is respectively 100\% and -100\% of portfolio value. \\
    constraint (\ref{last_carcon}) & $G =6$ & A portfolio can hold up to 6 different forward contracts. \\
    constraint (\ref{cashcon}) & $M = 10\%$ & Cash equivalent to 10\% of the value of each forwards position must be held in a portfolio. \\
    \bottomrule
    \end{tabular}%
  \label{tab:constr_tab}%
  }
\end{table}%

To compare risk and return of optimal portfolios with different constraints, an efficient frontier (Pareto front) is rendered accordingly. Comparison among frontiers is possible by visualisation. The frontier that is more north-western is considered more efficient as with the same level of risk, it offers higher return. An efficient frontier is constructed by setting a return target (constraint (\ref{port_ret})), then minimises the optimisation problem (\ref{MVproblem}) for an optimal portfolio and proceed to the next return target. We start from the monthly return level of 0.5\% to the maximum of 1.8\%. The return increment of each return level is set to 0.01\% which results in total 130 portfolios on each efficient frontier. 

All the experiments are run on PC (8GB RAM, CPU 2.10GHz) using \texttt{cplexmiqpex} package from CPLEX on MATLAB as a solver for mixed-integer quadratic programming. It is worth noting that although the optimisation model (\ref{MVproblem}) is originally convex, incorporating fixed and linear transaction costs of forward contracts eventually results in non-convex programming, \citep{Lobo2007}.

\subsubsection{Different Approaches to Implement Currency Overlay}
This analysis aims to study the effects of two strategies of currency overlay implementation on risk return reward of corresponding portfolios. Since currency overlay is devised to allow currency exposure to deviate from asset exposure to achieve better risk-return compensation and more efficient portfolios, there are basically two ideas to implement currency overlay.

\begin{itemize}
\item The unified approach -- optimises asset and overlay positions in a portfolio simultaneously for optimal asset and currency exposures. 
\item The two-stage approach -- begins with optimising portfolio allocation with no currency overlay and make an adjustment later. At the first step, exposure in asset and currency of investment in each country will be equal. The second step is primarily optimising currency exposure to make adjustments on original currency positions. This strategy, despite being a sub-optimal comparing to the unified approach, is practical for fine-tuning final currency exposure of existing portfolios. It could save significant transaction costs on hedging currency risk as all the alteration is made by entering new foreign exchange forward contracts while asset allocation remains unchanged.
\end{itemize}

\begin{figure}[H]
\begin{centering}
\includegraphics[scale=0.55]{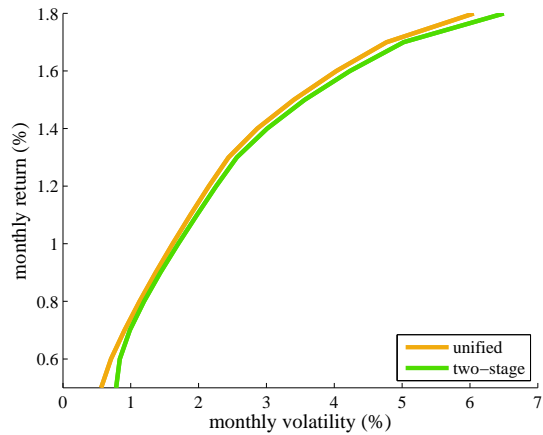}
\par\end{centering}
\caption{Efficient frontiers of portfolios optimised by unified and two-stage strategies. The unified approach optimises currency overlay and asset allocation at the same time while the two-stage approach optimises asset allocation first and then optimises the currency overlay on top of existing asset allocation to adjust the currency exposure. For currency overlay constraints, the key parameters are given as $V_u=100\%$ (unrestricted total overlay), $G =6$ (unrestricted number of forwards for currency overlay construction) and $M = 10\%$ (margin requirement for forwards is 10\%).}
\label{Exp1_1}
\end{figure}

The constraint parameters of each currency overlay implementation approach are set as in Table \ref{tab:constr_tab}. The resulting efficient frontiers are exhibited in Figure \ref{Exp1_1}. As expected, the two-step optimisation generates inferior risk-return rewards for portfolios. For each portfolio with equal volatility, the differences in portfolio returns obtained from the two approaches can be narrowed if there are further adjustments made on asset allocations.

\begin{figure}[H]
\begin{centering}
\includegraphics[scale=0.35]{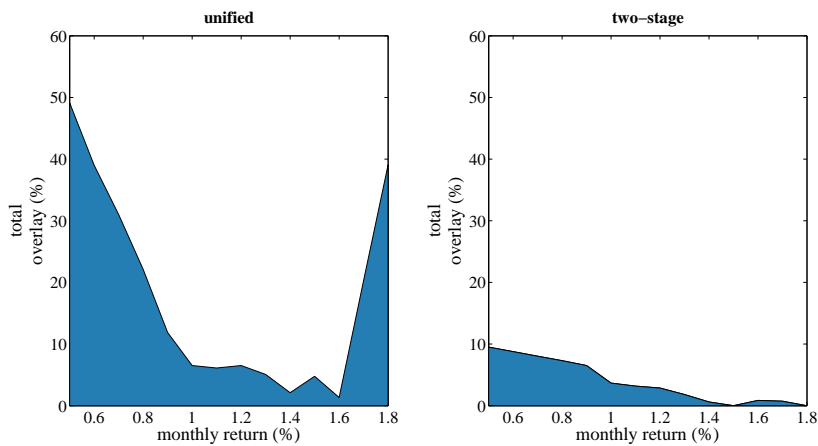}
\par\end{centering}
\caption{Total currency overlay, $V$, of portfolios optimised by the unified and two-stage strategies. For both cases, the key parameters controlling currency overlay constraints are $V_u=100\%$ (unrestricted total overlay), $G=6$ (unrestricted number of forwards for currency overlay construction) and $M=10\%$ (margin requirement for forwards is 10\%).}
\label{Exp1_2}
\end{figure}

The difference between two frontiers points out that modification on currency exposure alone is insufficient to significantly improve risk and return to portfolios. Larger deviation between frontiers is spotted in higher returns region. This is because high return portfolios tend to concentrate their investments in only one or two countries giving high pay-offs. Given that those asset allocations are fixed in the first place, then there is limited possibility to adjust original currency exposure to better position. This argument is supported by Figure \ref{Exp1_2}.

Currency adjustments on two-stage optimised portfolios are clearly less than those of portfolios optimised by the unified strategy. Focusing on the currency overlay of portfolios with two-stage approach (right panel of Figure \ref{Exp1_2}), it is observed that the overlay positions gradually decrease while portfolio returns increase. This is because forward contract exposures are not allowed to optimise along with asset allocation, thus at the first-stage the portfolios need to invest in stocks only in order to achieve high returns. Consequently, given that currencies generate lower risk and lower return comparing to stocks, the portfolios thus need no additional currency exposure otherwise the return target could not be satisfied. 

\subsubsection{Effects of Currency Hedging on Risk and Return of Portfolios}

The aim of this experiment is to emphasise the benefit of currency hedging on risk-return profile of portfolios. As described earlier, currency overlay is introduced to enable a portfolio to manage asset and currency exposures separately. In one extreme, a portfolio can be fully-hedged to have only USD exposure or, in another extreme, the portfolio is made fully-exposed to only foreign currencies. A balance between these twos provides a mix of exposures on local and foreign currencies which is expected to deliver the best risk-return profile to portfolios. 

We compare the results from three hedging policies. The first one is the `fully-hedged' policy where foreign exchange forwards are employed to remove entire foreign currency exposures (EUR, GBP and JPY) and remain only the exposure in USD (base currency of the portfolio). Referring to Table \ref{tab:port_structure}, this is achieved by enforcing $a_{11}+a_{21}+q_1+q_2+q_3 = 1$. The second policy is the `foreign' which uses forwards to remove USD exposure to foreign currencies, in this case the sum of foreign currency exposure is 100\% while USD exposure is 0\%, i.e., $a_{11}+a_{21}+q_1+q_2+q_3 = 0$. The last hedging policy is the `unrestricted' which makes no restriction on specific currency exposure. Noe that, in all the hedging policies presented, no constraint is imposed on asset exposure, so the portfolio can hold assets in Germany, UK and Japan while exposing to only USD exposure in the fully-hedged case. Other than the restriction on currency exposure, other constraints are as given in Table \ref{tab:constr_tab}.

\begin{figure}[H]
\begin{centering}
\includegraphics[scale=0.55]{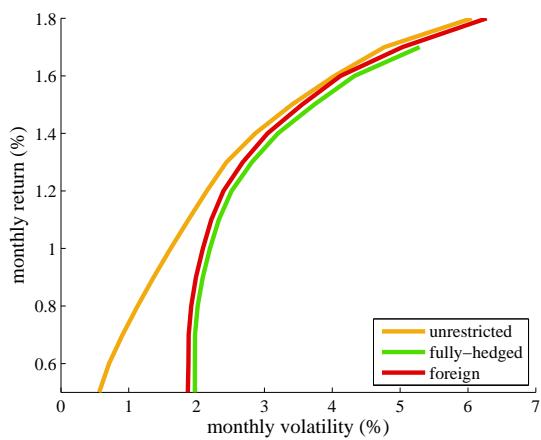}
\par\end{centering}
\caption{Efficient frontiers from different hedging policies. The `fully-hedged' restricts portfolios to have only currency exposure in USD. The `foreign' allows portfolios to expose only to foreign currencies. The `unrestricted' permits portfolios to expose to any currency. For portfolios under the `fully-hedged' policy, the highest return achievable is around 1.7\% monthly because foreign exchange gain from foreign currencies is not allowed. The efficient frontiers produced are subject to specific constraint according to each hedging policy. Apart from that, other constraints parameters are $V_u=100\%$ (unrestricted total overlay), $G=6$ (unrestricted number of forwards for currency overlay construction) and $M=10\%$ (margin requirement for forwards is 10\%).}
\label{Exp5_1}
\end{figure}

The resulting efficient frontiers from the experiment are displayed in Figure \ref{Exp5_1}. It can be seen that portfolios under the `fully-hedged' and `foreign' policies are clearly dominated by those from the unrestricted currency exposure policy, particularly at returns under 1.4\% monthly. 

Referring to the adjusted expected returns in Table \ref{tab:adjret}, foreign currencies supply higher return than the local currency (USD), hence limiting portfolios to only expose to USD dismisses an opportunity to reach high returns (in particular, approximately over 1.7\% monthly). The advantage of holding foreign currencies exposure also contributes in more efficient portfolios comparing to those holding only exposure in USD. However, the absence of USD means that portfolios are obligated to take more risk unnecessarily in some situations. For instance, at low returns where portfolios are supposed to hold more USD to preserve low volatility, disallowing USD exposure hence dramatically elevates higher volatilities at the lower end of the only-foreign-currencies efficient frontier.

\begin{figure}[H]
\begin{centering}
\includegraphics[width=12.5cm]{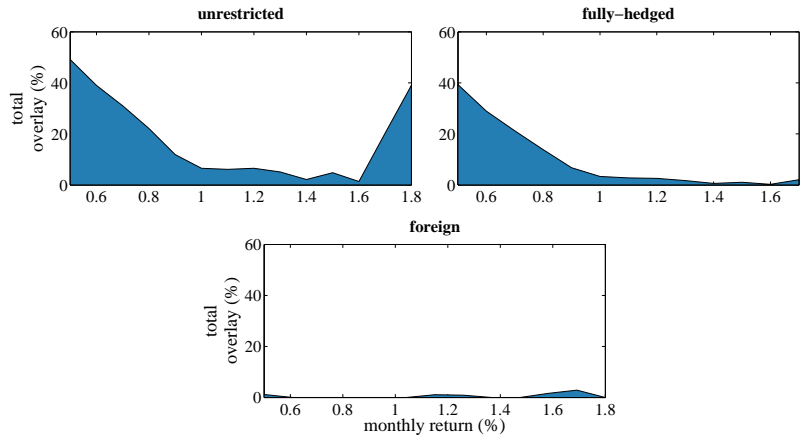}
\par\end{centering}
\caption{Total currency overlay, $V$, at each return level. Each panel represents each hedging policy. The total currency overlays in the `foreign' case are sparse as portfolios invest only in foreign countries. The fully-hedged portfolios require some currency overlay at low returns to remove foreign currency exposures from asset investments in German, UK or Japan while at higher returns, the portfolios tend to invest majorly in US equities which hence require only slight currency overlay.}
\label{Exp5_2}
\end{figure} 

It is also observed that the efficient frontiers lie closer when portfolios approach high return targets (above 1.4\%). Generally, at high returns, portfolios tend to hold risky assets in one or two countries (as diversification is second to return). For the fully-hedged policy, high returns could be achieved by holding large portion of US equities and expose to only USD while for the `foreign' case, portfolios invest in German and UK stocks and expose to only corresponding currencies. Such actions result in comparable returns and volatilities and are less inclined to currency overlay as portfolios expose to only currencies of the countries invested as shown in Figure \ref{Exp5_2}. For the unrestricted case, freedom  of choosing currency exposure along with the currency overlay helps reduce volatilities and thus its resulting portfolios are the most efficient among all hedging policies.

\subsubsection{Effects of Forward Contract Margin Requirement}

In our study, the transaction costs associated to holding forward contracts are the fixed costs, bid-ask spreads and margin requirement. The fixed costs depend on the number of forward contract pairs and the variable costs depend on bid-ask spreads which are associated to both size and number of forwards. Margin requirement is also considered a transaction cost as it requires portfolio to reserve more cash (with zero return) to cover forward positions. 

In general, higher transaction cost is expected to shift efficient frontiers south-eastern and lower risk-return compensation of portfolios. However, different return levels require different additional currency exposure from foreign exchange forwards. Thus at some return levels, increasing transaction costs might not worsen the risk-return reward as much as others. This study aims to see which part of efficient frontier are most and least affected by the variation of margin requirements in particular. 

\begin{figure}[H]
\begin{centering}
\includegraphics[scale=0.55]{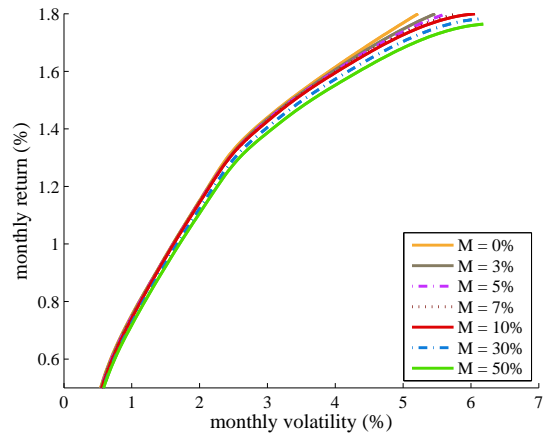}
\par\end{centering}
\caption{Efficient frontiers at different levels of margin requirement. $M$ determines how much cash is needed to be reserved for every 1\% of forward position on a portfolio. $M=0\%$ indicates that forward positions can be held without cash reserves while $M=5\%$ means that cash equivalent to 5\% of portfolio  is allocated to cash in order to cover a forward position of 1\% on a portfolio. Since the experiment particularly studies impact of different $M$, major overlay constraint parameters remain constant, i.e., $V_u=100\%$ (unrestricted total overlay) and $G=6$ (unrestricted number of forwards for currency overlay construction).}
\label{Exp2_1}
\end{figure}

We vary $M$ arbitrarily by 0\%, 3\%, 5\%, 7\%, 10\%, 30\% and 50\%. Other constraint parameters are set as in Table \ref{tab:constr_tab}. The efficient frontiers of optimal portfolios under different margin requirements exhibited in Figure \ref{Exp2_1} show that higher margins results in reduction of return per risk particularly when portfolio returns are high (greater than 1\% monthly approximately). 

\begin{figure}[H]
\begin{centering}
\includegraphics[scale=0.45]{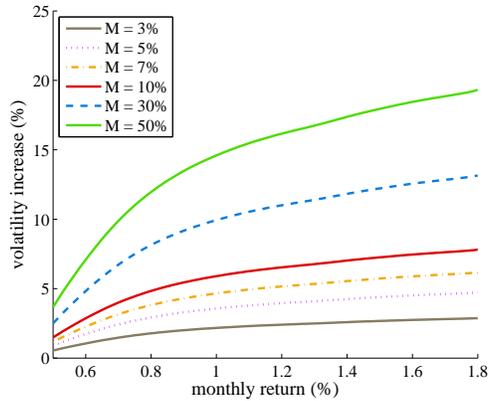}
\par\end{centering}
\caption{Percentage volatility increase relative to volatilities of the portfolio with no margin requirement ($M=0\%$). At each $M$, other key overlay constraint parameters are $V_u=100\%$ (unrestricted total overlay) and $G=6$ (unrestricted number of forwards for currency overlay construction).}
\label{Exp2_4}
\end{figure}

To highlight the effect of margin requirement levels, we plot the percentage increases of volatilities from the base case ($M=0\%$) at each return level in Figure \ref{Exp2_4}. The graph shows that a portfolio is more risky when margin requirement is higher since the advantage of holding forward contracts is offset by cash portion in the portfolio. This also implies that total currency overlay should be lower when margin requirement increases.  

\begin{figure}[H]
\centering
\subfloat{
        \label{subfig:correct}
        \includegraphics[scale=0.35]{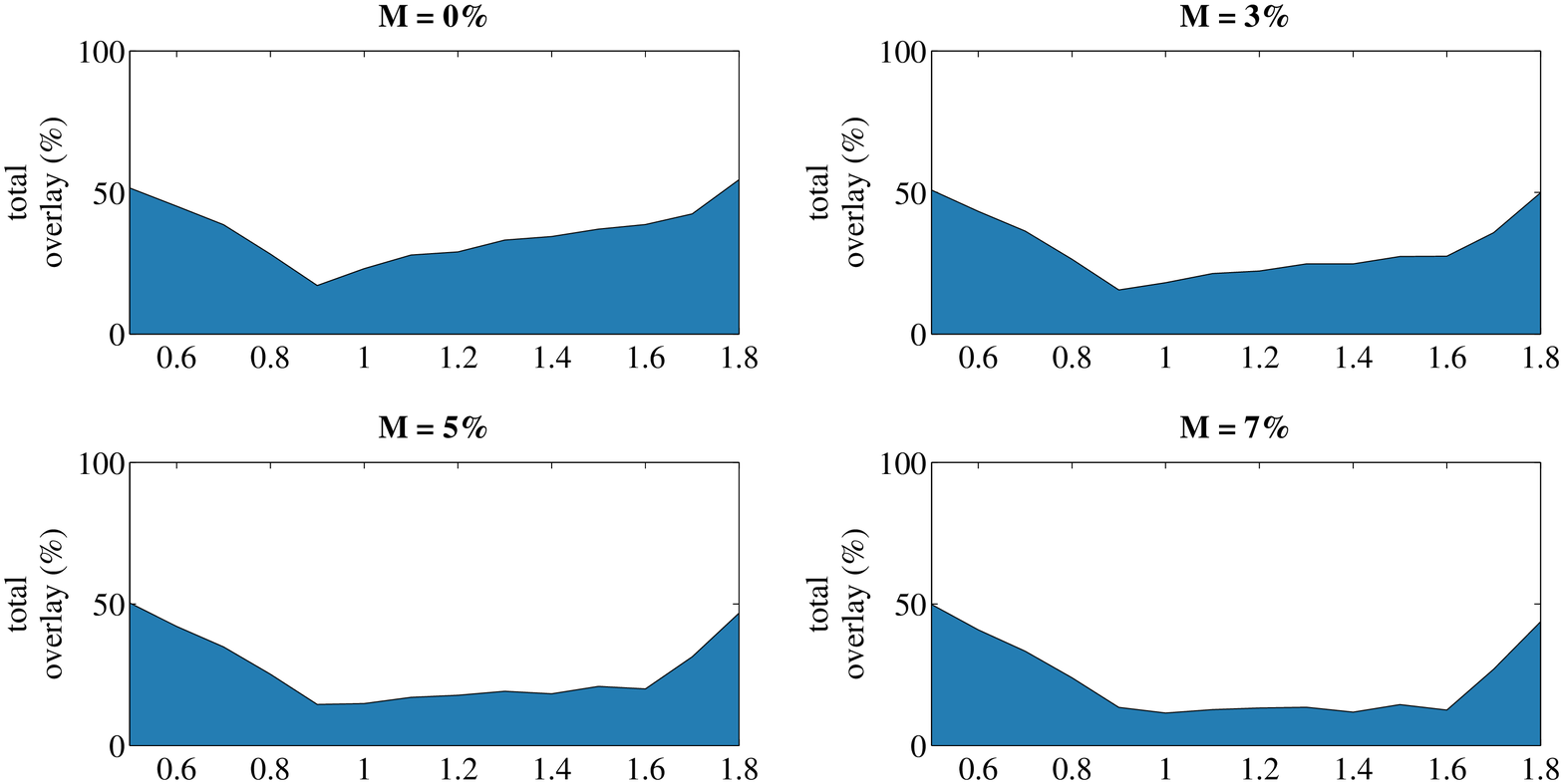}}
        \hfill
        \vspace{-0.75cm}
\subfloat{
        \label{subfig:notwhitelight}
        \includegraphics[scale=0.35]{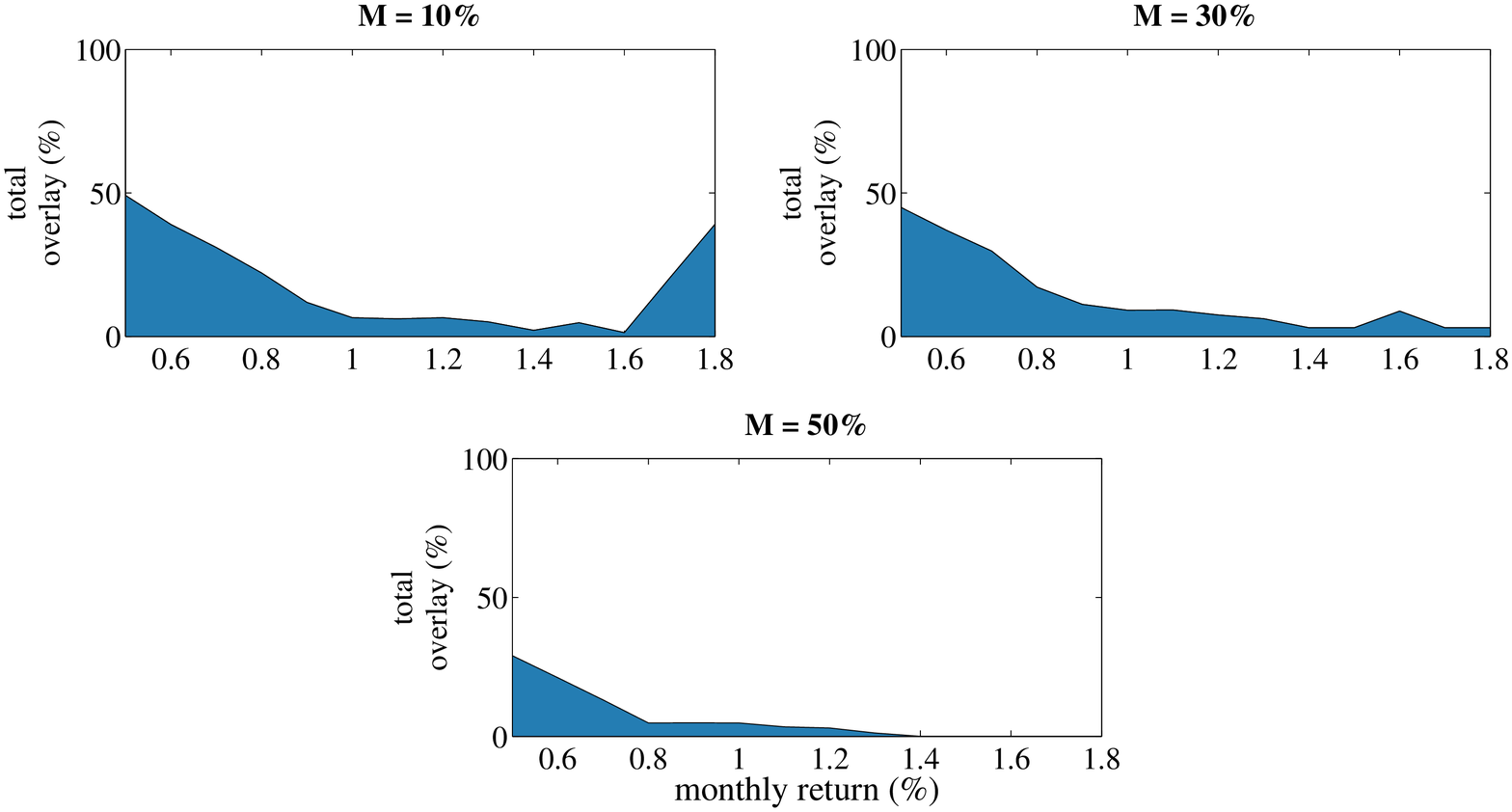}}
\caption{Total currency overlay, $V$, at each return level. Each panel represents different margin requirement indicated by values of $M$ over each panel. At each panel, other overlay constraint parameters are kept constant; $V_u=100\%$ (unrestricted total overlay) and $G=6$ (unrestricted number of forwards for currency overlay construction). At high returns, portfolios hold forwards in place of equities as they provide comparable returns with lower volatilities. However, such benefit of forwards diminishes when $M$ is large as returns from holding forwards are offset by large cash position which provides zero return. A large drop of total currency overlay at high returns is thus observed when $M$ increases.}
\label{Exp2_3}
\end{figure}

Figure \ref{Exp2_3} plots total currency overlays of portfolios at each level of margin requirement. When there is no margin requirement ($M=0\%$), no cash is needed in a portfolio, the efficient frontier under this condition is hence the most effective comparing to the others. 

Basically, holding forward contracts generates extra exposure on desired currencies without any purchase of physical assets. Given that currencies have comparable return and bear lower volatilities than equities (see Figure \ref{RiskReturn}), forwards are more favourable than stocks to raise portfolio returns. However, the benefit of competitive returns of forward contracts are offset with zero return on cash when margin requirement comes into effect. When maintenance margin of holding forward contracts grows larger, the attractiveness of forwards gradually diminishes as portfolios need to reserve more cash (with zero return) for forward positions. Higher margin requirement simply reduces return on holding forward contracts and leave them only beneficial for risk reduction. Therefore, when portfolio requires higher return, portfolios tend to give up forward contracts and hold risky assets instead.

It can be concluded that margin requirement of forward contracts affects portfolios most at the high return regions as the losing benefit of forward contracts causes portfolios to hold more risky assets to earn higher returns. In contrast, at lower return targets where risk reduction is a top priority, forwards are still in use as part of volatility reduction making low returns least affected. 

In following experiments, we try different approaches to reduce transaction costs occur from holding foreign exchange forwards. The first approach limits total currency overlay and another approach limits the number of forward contracts to construct currency overlay.

\subsubsection{Limiting Total Currency Overlay}
\label{exp_limitV}

The total currency overlay limit $V_u$ corresponds to the constraint (\ref{totovl}) which limits how much total currency exposure can be deviated from asset exposure by holding foreign exchange forward contracts. If $V_u$ is set as 0\%, currency exposure and asset exposure in each country are identical while $V_u = 1\%$ implies full flexibility of deviation in asset and currency exposures thereby the best option to improve risk-return reward to portfolios.

Confining total currency overlay limits forward exposure on portfolios which reduces transaction costs from bid-ask spreads and margin requirement. Since portfolios adjust currency exposure differently at each return level, the impact of limiting $V_u$ could also be dissimilar at different return targets. 

\begin{figure}[H]
\begin{centering}
\includegraphics[scale=0.55]{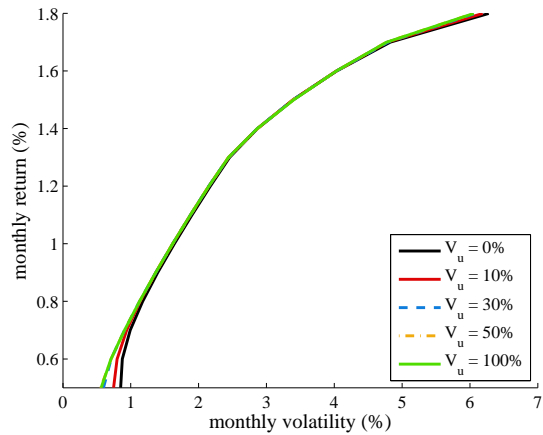}
\par\end{centering}
\caption{Efficient frontiers of optimal portfolios at different restriction of total currency overlay $V_u$. Larger $V_u$ allows flexibility for currency exposure to deviate from asset exposure, implying larger positions of forwards and higher transaction costs. Restricting $V_u$ hence reduces transaction costs associated to forwards at the expense of less efficient portfolios. Each efficient frontier is generated with different values of $V_u$ while other overlay constraint parameters remain constant; $G =6$ (unrestricted number of forwards for currency overlay construction) and $M = 10\%$ (margin requirement for forwards is 10\%). The frontiers when $V_u=50\%$ and $V_u=100\%$ coincide as total currency overlays never exceed 50\%.}
\label{Exp3_1}
\end{figure}

\begin{figure}[H]
\begin{centering}
\includegraphics[width=12.5cm]{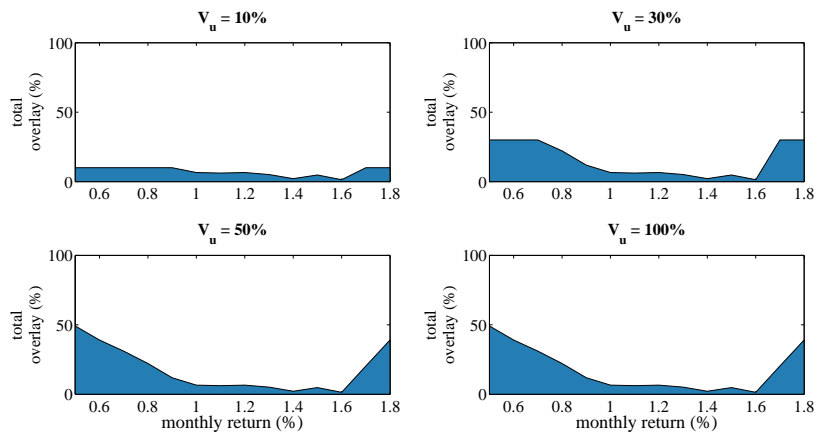}
\par\end{centering}
\caption{Total currency overlay, $V$, at each return level. Each panel represents different total overlay limit indicated by values of $V_u$ over each panel. Other currency overlay constraint parameters are $G =6$ (unrestricted number of forwards for currency overlay construction) and $M = 10\%$ (margin requirement for forwards is 10\%). The plot of $V_u=0\%$ is not included in the figure as it presents no currency overlay.}
\label{Exp3_2}
\end{figure}

In the experiment we compare the optimal portfolios from five arbitrary values of $V_u$ which are 0\%, 10\%, 30\%, 50\% and 100\%. Other constraints are applied as in Table \ref{tab:constr_tab}. Figure \ref{Exp3_1} exhibits efficient frontiers obtained from optimisation. The one produced by setting $V_u=0\%$ demonstrates lowest risk-return compensation while the others generated from higher values of $V_u$ show improvement on the risk-return trade-off. It is observed that in all settings of $V_u$, total currency overlays $V$ on the portfolios never exceed 50\% (see Figure \ref{Exp3_2}), hence optimal portfolios obtained from $V_u=50\%$ and $V_u=100\%$ are identical and their efficient frontiers coincide.

\begin{figure}[H]
\begin{centering}
\includegraphics[scale=0.55]{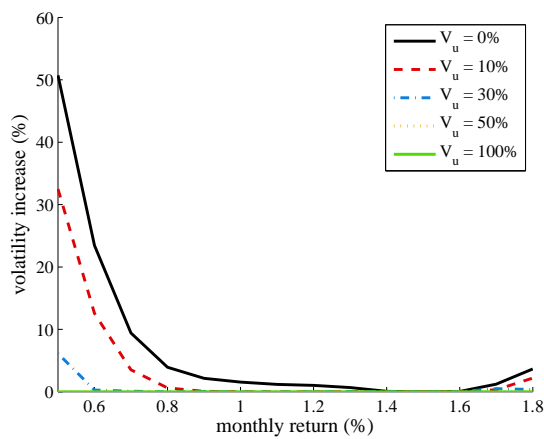}
\par\end{centering}
\caption{Percentage volatility increase relative to volatility of portfolios with no restriction on currency overlay ($V_u=100\%$). For each $V_u$, portfolios are optimised under currency overlay constraints with following parameters, $G =6$ (unrestricted number of forwards for currency overlay construction) and $M = 10\%$ (margin requirement for forwards is 10\%). Since $V_u=50\%$ and $V_u=100\%$ produce identical optimal portfolios, the relative volatility increase for $V_u=50\%$ is 0.}
\label{Exp3_3}
\end{figure}

Figure \ref{Exp3_3} shows percentage volatility increase relative to the case of $V_u=100\%$ at each return level. It appears that at low returns, portfolio volatilities increase most when total currency overlays are bounded and this effect is lessened when portfolio return increases. The least affected regions are when returns are approximately between 1-1.6\% monthly where portfolios need only modest adjustments on currency exposure to achieve optimal portfolios.

Figure \ref{Exp3_2} illustrates the total overlay at each return level of a portfolio. Since $V_u=0\%$ indicates currency exposure and asset exposure are identical, there presents no overlay position. When $V_u=10\%$ and $V_u=30\%$, the constraint (\ref{totovl}) is binding suggesting that the risk-return profile can be further improved if the total overlay limit is loosened and when $V_u=50\%$ and $V_u=100\%$, total currency overlay usages are identical as portfolios never require total currency exposure adjustment over 50\% of portfolio values.

\begin{figure}[H]
\begin{centering}
\includegraphics[width=12.5cm]{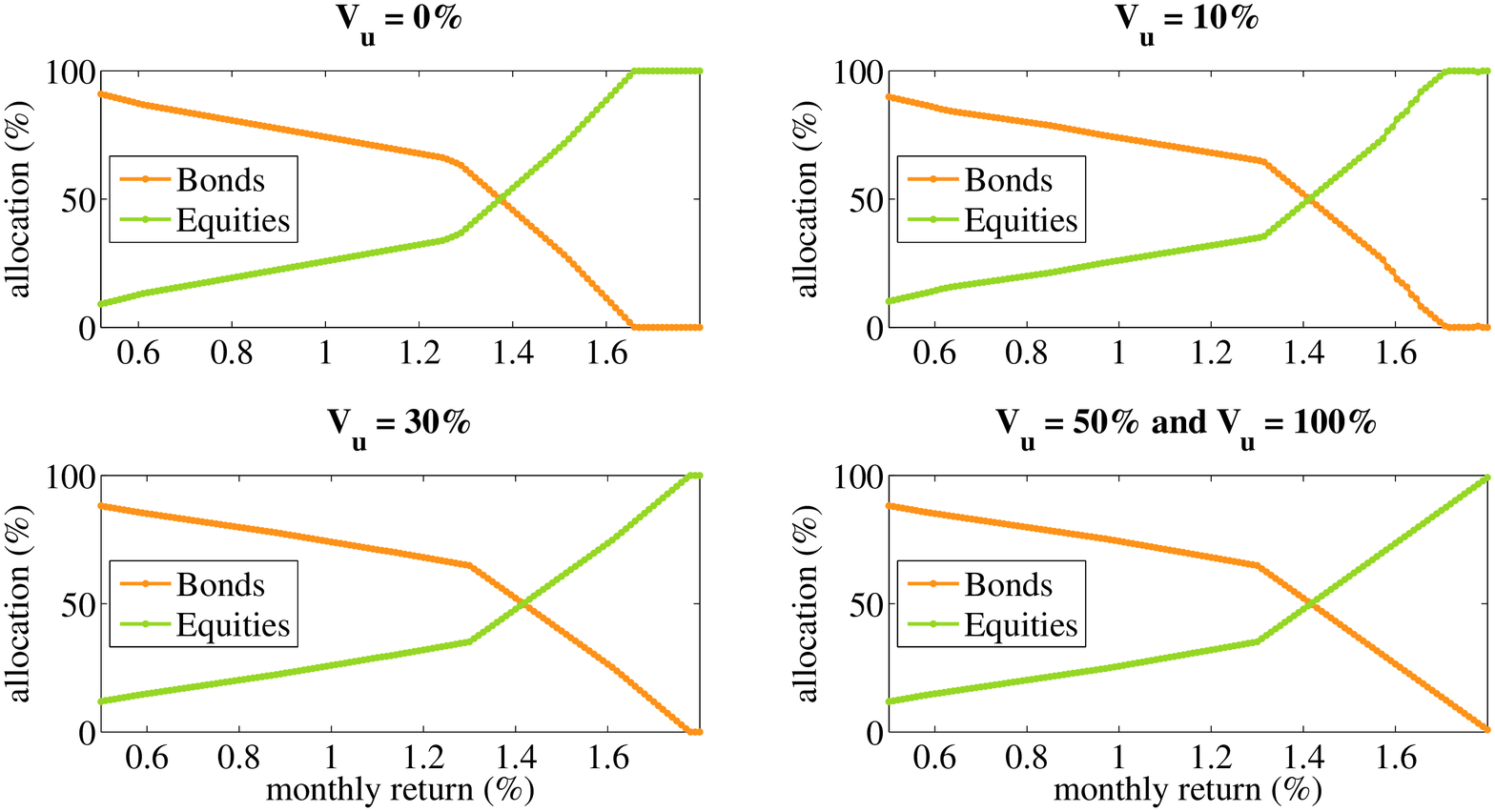}
\par\end{centering}
\caption{Composition of total bond and total equity holdings at at each return level. Each panel represents different total currency overlay limit $V_u$. At each panel, apart from the varied $V_u$, other constraint parameters are fixed as $G =6$ (unrestricted number of forwards for currency overlay construction) and $M = 10\%$ (margin requirement for forwards is 10\%).}
\label{Exp3_4}
\end{figure}

Generally, a portfolio gains higher return by reducing allocation of low-risk investments for riskier investments. Figure \ref{Exp3_4} demonstrates aggregate holdings in bonds and equities across countries at different values of $V_u$. It shows that bond allocations gradually decrease along with increasing returns. However, the decrease in bond allocation is slower when $V_u$ is more relaxed. This signifies that additional currency exposure from overlay positions contributes the holding of risky investments in tandem with equities. 

The plot of volatilities and returns in Figure \ref{RiskReturn} shows that EURUSD gives comparable return to equities while having significantly lower volatility. Holding substantial portion of EUR-cross forward contracts along with equities assists risk reduction of portfolio rather than investing in stocks alone and results in better risk-return trade-off for portfolios.

\subsubsection{Limiting Number of Forward Contracts} 

\label{exp_limitG}
For an investment in four currencies, currency overlay can be constructed from up to six different foreign exchange forwards which are shown in Table \ref{tab:port_structure}. In this experiment, referring to the constraint (\ref{last_carcon}), we vary values of $G$ from 1 to 6 to limit the number of forward contracts spent in overlay construction. Fewer number of forward contracts cut down the fixed operating costs along with variable costs from bid-ask spreads and margin requirement. This is opposed to limiting the total currency overlay in which transaction costs are lessened primarily from reduced exposure while the fixed costs might remain the same.

In terms of computation, values of $G$ between 1 and 5 create a cardinality constraint on the selection of forward pairs constituting currency overlay. For example, $G=3$ forces a portfolio to choose only three or fewer forwards from Table \ref{tab:port_structure} that can minimise variance of the portfolio at a given return level. The other constraint parameters are applied as in Table \ref{tab:constr_tab}.

\begin{figure}[H]
\begin{centering}
\includegraphics[scale=0.55]{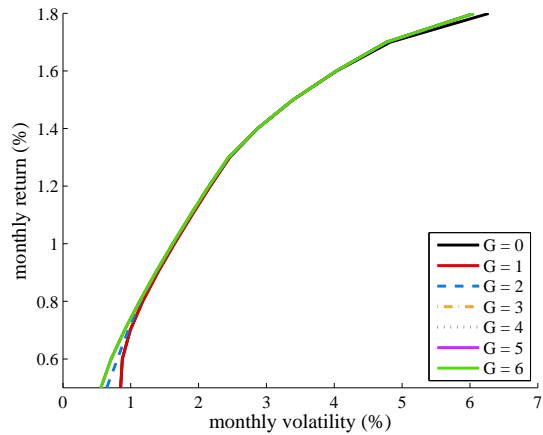}
\par\end{centering}
\caption{Efficient frontiers produced under different limits on the number of forward contracts allowed. $G$ represents for the maximum number of forward contracts allowed to construct currency overlay. Fewer contracts reduce the fixed operating cost but also impair risk-return profile of portfolios. For the case of four currencies, $G=0$ implies currency overlay is not permitted while $G=6$ implies all possible forward contracts can be used to create currency overlay. At each $G$, parameters of other currency overlay constraints are $V_u=100\%$ (unrestricted total overlay) and $M = 10\%$ (margin requirement for forwards is 10\%).}
\label{Exp4_1}
\end{figure}

\begin{figure}[H]
\begin{centering}
\includegraphics[scale=0.35]{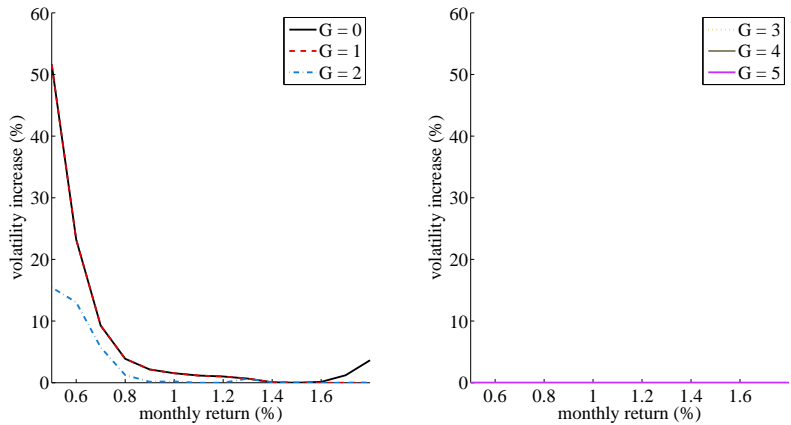}
\par\end{centering}
\caption{Percentage volatility increase relative to volatilities of the portfolio that allows up to 6 forward contracts to construct currency overlay ($G=6$). At each $G$, parameters of other currency overlay constraints are set as $V_u=100\%$ (unrestricted total overlay) and $M = 10\%$ (margin requirement for forwards is 10\%). For the case of 4 currencies, only 3 forward contracts are sufficient to create all overlay positions. Imposing $G\geq3$ (right panel) therefore produces similar portfolios and their relative volatility increases from the case of $G=6$ are zero.}
\label{Exp4_3}
\end{figure}

In Figure \ref{Exp4_1}, the significant difference of efficient frontiers is visible when increasing $G$ from 0 to 1 and 2 in which the efficient frontiers shift north-west when $G$ increases. Figure \ref{Exp4_3} displays percentage volatility increase relative to the most relaxed case ($G=6$). It is obvious that more freedom to choose forward contracts to construct currency overlay produces better return per risk ratio although the increment of improvement is less noticeable when $G \geq 3 $ than when $G$ increases from 0 to 2. In addition, returns lower than 1\% monthly experience larger volatility increase than higher returns as flexibility to modify currency exposure is diminished with limited number of available forward contracts.

When stepping from $G=0$ to $G=1$, the risk-return reward is improved as portfolios are allowed to hold forward contracts rather than stocks thereby lowering total volatility. Stepping from $G=1$ to $G=2$ is a substantial progress as $G=1$ permits an adjustment of currency exposure on only two out of four currencies while $G=2$ can cover all four currencies (with USDEUR and GBPJPY) but with limited flexibility\footnote{With two forwards, only two pairs of currency exposure can be modified. For example, with USDEUR and GBPJPY, exposures on EUR and GBP depend on USD and JPY, respectively and cannot be altered separately until more number of forward contracts are allowed.} to adjust exposure of each currency. $G=3$ makes further flexibility to modify currency exposure as some patterns of currency overlays cannot be produced by two forward contracts. For instance, the following overlay is impossible to be replicated by two forward contracts but possible with three or more.

\begin{table}[H]
  \small
  \centering
  \caption{Example of currency overlay which needs three or more forward contracts to produce.}
    \begin{tabular}{rcccc}
    \toprule
          & USD   & EUR   & GBP   & JPY \\
    \midrule
    overlay (\%) & 1     & 5     & -2    & -4 \\
    \bottomrule
    \end{tabular}%
\end{table}%

Thus shifting from $G=2$ to $G=3$ or more significantly provides improvement to risk-return reward of portfolios. In fact, for $C$ currencies, there needs $C-1$ different forward pairs to cover all possible currency overlays. Therefore allowing $G\geq3$ generates the same optimal portfolios.

\begin{figure}[H]
\begin{centering}
\includegraphics[width=12.5cm]{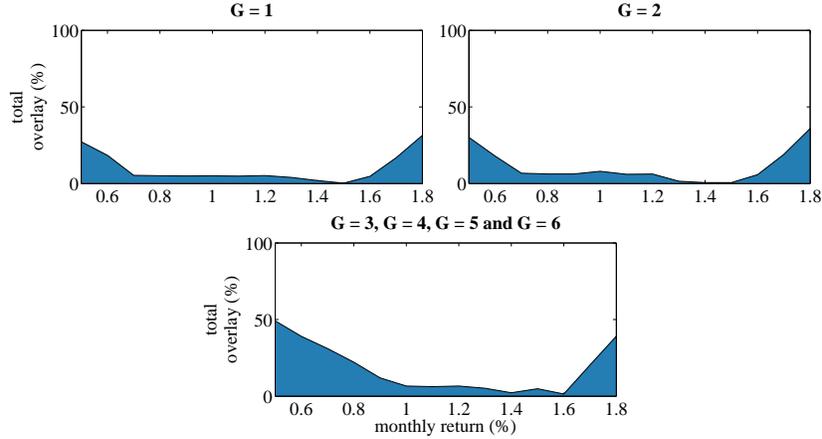}
\par\end{centering}
\caption{Total currency overlay at each return level. Each panel represents maximum number of forward contracts used in currency overlay. At each $G$, other currency overlay constraint parameters are $V_u=100\%$ (unrestricted total overlay) and $M = 10\%$ (margin requirement for forwards is 10\%). The plot of $G=0$ is not included in the figure as there is no forward contract allowed and hence no currency overlay. For the case of four currencies, three forward contracts are sufficient to replicate all possible currency overlays, the resulting total overlays for $G\geq3$ are therefore similar.}
\label{Exp4_2}
\end{figure} 

Figure \ref{Exp4_2} shows total currency overlay at each return target. Comparing the top panel ($G=1$ and $G=2$) and bottom panel ($G\geq3$), we notice a large drop of total currency overlay when returns are less than 1\% monthly while only a slight decrease in total overlay position is observed at higher returns. These results coincide with relative volatility increase in Figure \ref{Exp4_3}. 
It can be explained that at high returns, currency overlays are generally composed of one or two forward contracts associated to EUR or GBP to raise target returns. 
On the other hand, at low returns and low risks, more forward contracts are needed in order to reduce volatility. For example, in order to hedge risk from all foreign currencies, the pairs USDEUR, USDGBP and USDJPY are needed to modify currency exposure to solely USD. Thus saving transaction cost by restricting the number of forward contracts affects portfolios most when target returns are low (1\% monthly with our data).

Besides controlling transaction costs associated to forward contracts, another advantage of imposing cardinality constraint also shortens the runtime. Although the cardinality constraint transforms the optimisation problem to mixed-integer programming but the search space is reduced at the same time. 

\begin{table}[H]
\centering
\begin{tabular}{cc}
    \begin{minipage}{.4\linewidth}
    \small
    \centering  
    \caption{
		Comparison of average runtime taken on optimising each portfolio in the two experiments: limiting number of forward contracts (left) and limiting total currency overlay (right).
		For the left panel, fixed parameters are $V_u=100$ (unrestricted total overlay) and $M = 10\%$ while $G$ is varied. Restriction on the the maximum number of forward contracts $G$ is considered a cardinality constraint which narrows down the search space and reduces the runtime accordingly. 
		For the right panel, constant currency constraint parameters are $G=6$ and $M = 10\%$ (margin requirement for forwards is 10\%) while $V_u$ is varied. 
		The number of decision variables in the experiments are 14 real variables for both panels, with another 6 binary variables for the cardinality constraints in the left panel.
	}  
    \begin{tabular}{cc}
    \toprule
    $G$ varies, $V_u$ fixed       & cpu time \\
    (experiment \ref{exp_limitV}) & in seconds \\
    \midrule
    0     & 1.7 \\
    1     & 1.8 \\
    2     & 1.8 \\
    3     & 1.9 \\
    4     & 2.2 \\
    5     & 2.3 \\
    6     & 2.5 \\
    \bottomrule
    \end{tabular}%
  \label{tab:runtime}%
    \end{minipage} &

    \begin{minipage}{.4\linewidth}
    \small
    \centering    
    \begin{tabular}{cc}
    \toprule
    $V_u$ varies, $G$ fixed       & cpu time \\
    (experiment \ref{exp_limitG}) & in seconds \\
    \midrule
    0.0   & 1.7 \\
    0.1   & 2.3 \\
    0.3   & 2.3 \\
    0.5   & 2.4 \\
    1.0   & 2.3 \\
          & \\
          & \\
    \bottomrule
    \end{tabular}%
    \end{minipage} 
\end{tabular}
\end{table}%

Table \ref{tab:runtime} presents time taken on running each portfolio in the two experiments, limiting total currency overlay and limiting number of forward contracts. 
The number of decision variables for the experiment limiting the number of forwards (left panel) is 20 (including binary decision variables) and it is 14 for the experiment varying $V_u$ (right panel). 
Smaller values of $G$ reduce the search space and hence decrease the runtime while when $G$ is fixed (in the experiments in Section~ \ref{exp_limitG}) there is insignificant difference of runtime over variation of $V_u$. 
That is, in our study, although the problem is non-convex due to the inclusion of fixed and variable transaction costs of foreign exchange forwards, all the problems are still solvable with CPLEX with inexpensive runtime. 
However, given that the number of decision variables so small, although the runtimes of Figure~\ref{tab:runtime} are low it might well be that heuristic or metaheuristic methods would be able to find good answers even more quickly, and this may be useful for the future extensions discussed in the next section.

\clearpage
\section{Conclusions}  \label{sec:concl}

Currency overlay is defined as deviation between asset and currency exposures in a portfolio. In this study, currency overlay is created by holding foreign exchange forward contracts. 
We introduce a novel portfolio optimisation model that incorporates currency overlay to allow flexibility in investing asset and exchange rate in each country. 
The key feature is that overlay positions are designed to be structured by different forward contract pairs which allow us to impose different transaction costs on each forwards pair and implement practical constraints to reduce overall transaction costs. 
Besides, we introduce the approach to incorporate the cost of carry of forward contracts into the process of return and risk calculation of the portfolios which reflects more accurate gain and loss of forward contracts held by portfolios and helps complete calculating total return and volatility.

We presented four experiments to investigate effects on risk-return compensation of portfolios when different types of constraints are imposed. 
The first experiment is to study two different strategies in implementing currency overlay, the unified and the two-stage approaches. 
The results support that allowing portfolios to optimise currency overlay and asset allocation at the same time yields better risk-return reward than optimising asset allocation first and adjusting the currency exposure afterwards. 
The second experiment shows that a fully-hedged portfolio or a portfolio that exposes to all foreign currencies are not as efficient as a portfolio whose currency exposure is from local and foreign currencies combined. 
The third experiment concludes that large margin requirement diminishes benefit of holding currency overlay and induces portfolios to hold only risky assets and ignore forward contracts in order to achieve high returns (above 1\% monthly approximately). Subsequently, the paper focuses on imposing constraints to reduce transaction costs occurred from holding currency overlay. 
The fourth experiment enforces the limits of total currency overlay which directly reduces costs from bid-ask spread and opportunity lost from margin requirement. 
The results show that the more total overlay is allowed, the better the risk-return compensation on portfolios. Besides, the findings show that portfolios utilise more currency overlay when target returns are lower than 1\%. 
These low returns regions therefore experience significantly high relative volatility increase comparing to the cases when portfolio returns are higher. 
The last experiment limits the maximum number of forward contracts to construct currency overlay with the aim to reduce the fixed operating cost per forward contract. 
The experiment results show that for portfolios investing in four currencies, three different forward contract pairs are sufficient to replicate all currency overlay positions. 
For the impact on risk-return tradeoff, the results are similar to the case of limiting total currency overlay. 
At returns lower than 1\%, portfolios have highest relative volatility increase as portfolios tend to hold more forward contracts to hedge currency risk at low returns while at higher returns, only few forward pairs that deliver high returns are required to achieve the return targets.

In all experiments, we notice different total currency overlay position at each return target. This indicate that portfolios adjust their currency exposures differently at each return target. 
Generally, currency overlays are large at both ends of efficient frontiers where returns are either low or high while smaller overlay positions are observed at moderate return targets. 
At low returns, portfolios require different forward contracts to hedge foreign currencies to keep low-risk-low-return profile. 
On the other hand, when portfolios require higher returns, forwards are held instead of equities as they provide competitive returns with lower volatilities. 
Such advantage of forwards, however, is dissolved if there exists margin requirement to sustain forward positions. 
This also suggests that counterparties that offer lower margin requirement are more favourable given that their credibility is not a concern. 
Lastly, all the experiment results signify that reducing forwards transaction costs through constraints imposition affects portfolios differently as overlay positions are varied by return targets. 
The middle-range returns where small currency exposure adjustment is required are therefore the least affected area on efficient frontiers.

\subsection{Future Work}

We intend that this work should be combined with existing work in portfolio optimisation and so extend such work to international portfolio optimisation. 
For instance, the portfolio can invest in more countries, currencies, asset classes and more importantly, invest in individual securities rather than indices. 
This should remove the limitation in current methods that bond or stock indices are constructed disregarding co-movement of asset across currencies; investing in individual assets allows full exploitation of cross-currency correlation. 

In this case, the number of different assets will become much larger, and then a natural step is to 
impose cardinality constraint on the number of individual assets on the portfolio.
There could well be separate cardinality constraints covering all of the assets and for each individual currency. 
Similarly, minimum holding position and minimum transaction lots are expected to be added for a more realistic optimisation problem; for instance in \citep{Soleimani2009,Lin2008}. In addition, robust solutions should be produced and that take account of the uncertainties in risk, return, and now also exchange rates; for example, \citep{Topaloglou2008,Fonseca2011}.

These extensions to a ``robust cardinality-constrained international portfolio optimisation with currency overlay'' problem will enlarge the problem scale, making an interesting, useful, and computationally challenging problem domain, intrinsically using the currency overlay methods presented in this paper.





\bibliographystyle{apalike}
\bibliography{Bibliography}







\end{document}